\documentclass{aa}  
\usepackage{graphicx}
\usepackage{txfonts}
\usepackage{epsfig}
\usepackage{natbib}

\newcommand{\bcep}{$\beta$ Cep}
\newcommand{\nvi}{N\,{\sc vi}}
\newcommand{\ovii}{O\,{\sc vii}}
\newcommand{\neix}{Ne\,{\sc ix}}
\newcommand{\refcom}{}
\newcommand{\xmm}{XMM-\emph{Newton}}

\begin{document}
\title{Testing \refcom{magnetically confined wind shock models} for $\beta$\,Cep using \xmm\ and \emph{Chandra} phase-resolved X-ray observations}
\author{F. Favata\inst{1}
\and C. Neiner\inst{2}
\and P. Testa\inst{3}
\and G. Hussain\inst{4}
\and J. Sanz-Forcada\inst{5}
%\fnmsep\thanks{}
}

\offprints{F. Favata}

\institute{European Space Agency, 8-10 rue Mario Nikis, 75015 Paris, France\\
\email{Fabio.Favata@esa.int}
\and GEPI, UMR 8111 du CNRS, 5 place Jules Janssen, 92195 Meudon Cedex, France
\and Kavlis Institute for Astrophysics and Space Research, MIT, Cambridge, Mass., USA
\and European Southern Observatory, Garching bei MŸnchen, Germany
\and Laboratorio de Astrof'sica Espacial y F'sica Fundamental, INTA, PO Box 50727, 28080 Madrid, Spain
}

\date{Received ; accepted }

% \abstract{}{}{}{}{} 
% 5 {} token are mandatory
 
\abstract
{}
{We have performed a set of high- and low-spectral resolution phase-resolved X-ray observations of the magnetic B star \bcep, for which theoretical models predict the presence of a confined wind emitting X-rays from stationary shocks. \refcom{Some of the} models predict, given the peculiar geometry of \bcep, strong rotational modulation of the X-ray emission, \refcom{while other models predict a much lower amplitude modulation at 90 deg phase shift from the modulation predicted from the first group of models}. Our observations were designed to provide a stringent test of such models.}
{We obtained four observations spaced in rotational phase with \xmm\ (using both the EPIC cameras and the RGS spectrograph) and with \emph{Chandra} (using the LETG spectrograph). A detailed analysis of the data was performed to derive both photometric and spectral parameters from the EPIC data, searching for rotational modulation, and to derive the location of the X-ray plasma from the line ratios in the He-like triplets of N, O and Ne from the RGS data. The LETG data were used to constrain the presence of bulk motions in the plasma. }
{The strong rotational modulation predicted by the \refcom{early, static} magnetically confined wind model for the X-ray emission is not observed in \bcep. The small modulation present goes in the opposite direction, pointing to the absence of any optically thick disk of neutral material, \refcom{and showing a modulation consistent with the later, dynamic models of magnetically confined wind models in B stars}. The lack of observed bulk motion points to the plasma being confined by a magnetic field, but the low plasma temperature and lack of any flaring show that the plasma is not heated by magnetic reconnection. Therefore, the observations point to X-ray emission from shocks in a magnetically confined wind, with no evidence of an optically thick, dense disk at the magnetic equator.}
{}

\keywords{}

\authorrunning{Favata et al.}
\titlerunning{X-ray emission from \bcep}

\maketitle

\section{Introduction}

Early-type stars were established as strong soft X-ray sources already during the very first survey of stellar X-ray emission conducted with the \emph{Einstein} observatory (\citealp{vcf+81}). O and B stars lack the external convection layer which is an essential component of the dynamo in late-type dwarfs, and thus are expected to lack the highly structured magnetic fields which confine and heat solar-type coronae. The observed X-ray emission is thermal in nature, and in general of lower temperature than observed in active late-type stars. The mechanism initially proposed to explain the observed X-ray emission was self shocking in the radiatively driven fast, strong stellar winds which are a characteristic of massive stars (\citealp{lw80}). This scenario makes a well-defined prediction, i.e.\ that emission lines should be broad and blue-shifted. The advent high spectral resolution made possible by the launch of \xmm\ and \emph{Chandra} has shown that some massive stars do indeed show the expected signature of X-ray emission from the wind (\refcom{e.g.\ $\zeta$~Pup,} \citealp{cmw+2001}). Other stars however show narrow spectral lines \refcom{(e.g.\ $\theta^1$ Ori C, \citealp{goc+2005}).} \refcom{Narrow lines imply a low-velocity plasma which has been interpreted as a confined plasma, likely requiring magnetic fields (e.g.\ \citealp{sch+2003}, \citealp{sth+2006} and reference therein). Alternative scenarios not requiring magnetic confinement for the X-ray emission from early-type stars showing narrow lines have also been put forward (\citealp{clg+2006}; \citealp{lpk+2006}). A recent survey of early-type high-resolution spectra has been performed by \cite{wc2007}.}

Recently, dipolar fields have been detected in a number of massive stars, in some cases with the magnetic axis at a significant angle from the rotation axis. Scenarios implying magnetic confinement of the wind from a dipolar magnetic field have been developed, and (as discussed below) they succeed in explaining several of the observed characteristics in massive stars with a measured magnetic fields and X-ray emission. For example, \cite{bm97a} developed a magnetically confined wind shock model (MCWS) to explain the characteristics of the X-ray emission from Ap and Bp stars, and \citet{bm97b} applied the MCWS model to explain some of the salient characteristics of the X-ray emission of   $\theta^1$ Ori C, one of the only two O-type stars with a detected magnetic field. $\theta^1$ Ori C shows strong modulation of the X-ray emission with the rotational period, with a nearly sinusoidal light curve and a $\simeq 50\%$ peak-to-peak modulation amplitude (\citealp{gcs+97}; \citealp{sfm+2005}); such strong modulation agrees very well with the predictions of the MCWS scenario. Interestingly, the other O star on which a magnetic field has recently been detected, HD 191612 (\citealp{dhb+2006}) has an unusually slow rotation rate, and it also has a field with a strong dipolar component. 

As discussed in Sect.~\ref{sec:bcep}, $\beta$\,Cep has a significant dipolar magnetic field \refcom{($\simeq 360$ G, \citealp{hjd+2000})}, with the magnetic axis at almost 90 deg from the rotational axis, and it therefore is a particularly interesting star for the study of the magnetically confined wind model. Its X-ray emission had already been detected by the \emph{Einstein} observatory, and \cite{dwb+2001} (hereafter D01) devised a detailed model for it, based on the MCWS scenario \refcom{of \citet{bm97b}} and the later ROSAT observations. \refcom{They make} clear and verifiable predictions on the temporal variability of the X-ray emission and on its spectrum, as well as on the spatial location of the X-ray plasma. 

\refcom{Later work based on dynamical modeling of the wind-magnetic field interaction however showed that the thick disk predicted by \citet{bm97b} would be unlikely to form around a star like \bcep\ (\citealp{goc+2005}; \citealp{uo2002}; \citealp{to2005}). }

We have carried out a campaign of X-ray observations using both \emph{Chandra} and \xmm, specifically designed to provide a stringent test of MCWS-based model of the X-ray emission of \bcep. The observations were designed to determine whether the X-ray emission presents the variability with rotational phase predicted by the D01 model, and (by using triplet ratios and Doppler shifts) at which distance from the star's photosphere the bulk of the X-ray emission is concentrated. Our interest in the X-ray emission of \bcep\ was initially also driven by its being a Be star, although, as discussed in Sect.~\ref{sec:bcep}, recent observations have suggested that the Be phenomenology is due to the secondary (presumably much less X-ray active) companion in the \bcep\ system, rather than to the magnetic, X-ray active primary.

The present paper is structured as follows: after the Introduction, the characteristics of our target star, \bcep, are discussed in Sect.~\ref{sec:bcep}. \xmm\ and \emph{Chandra} observations are discussed in Sect.~\ref{sec:obs} with their analysis, and the relative results are presented in Sect.~\ref{sec:res}. Finally, our conclusions are presented in Sect.~\ref{sec:concl}.

\section{The star $\beta$\,Cep}
\label{sec:bcep}

$\beta$\,Cep stars are early-B subgiants or giants that exhibit coherent short-period radial velocity variations, successfully explained in terms of pulsations with periods ranging from about 3 to 8 hours. The driving mechanism of pulsation in $\beta$\,Cep stars is the $\kappa$ mechanism, i.e.\ an effect of the changing opacity of iron-peak elements deep in the stellar envelope (\citealp{dp93b}). 

The prototype of this class, the star $\beta$\,Cep itself (HD\,205021, HR\,8238), long classified as a B1IIIe star, is at a Hipparcos distance of 182 pc. The literature values of the basic photospheric parameters of \bcep\ span a relatively wide range: $T_{\rm eff}$ varies from 24\,000 K (e.g.\ \citealp{hws94}) to 27\,000 K (e.g.\ \citealp{tmd+2003}), $\log g$ from 3.31 (e.g.\ \citealp{hws94}) to 4.07 (e.g.\ \citealp{cb92}), while $M$ ranges from $9.9 M_{\odot}$ (D01) to $16.4 M_{\odot}$ (\citealp{hws94}). The radius $R$ has been estimated from $6.5\, R_{\odot}$ (D01) to $8.6 R_{\odot}$ (\citealp{hh77}). The main pulsation period of $\beta$\,Cep, 4 hours and 34 minutes, corresponds to a radial $p$ mode, but additional periods exist due to non-radial modes. While \cite{dn2005} and \cite{nd2005} derive a photospheric metal abundance very close to solar ($m/H = -0.07 \pm 0.10$) from the analysis of IUE spectra, \citet{mba+2006} recently performed a detailed abundance analysis of the photosphere of \bcep\ based on high-resolution optical spectra, deriving abundances significantly lower than solar values, except for N and possibly S, as detailed in Table~\ref{tab:abun}. In the following we adopt the abundances of \citet{mba+2006}.

\begin{table}[htdp]
\caption{The photospheric abundances of \bcep\ as determined by \citet{mba+2006}. The first column gives the ratio between the photospheric abundance of \bcep\ and the standard photospheric abundance of the Sun of \citet{gs98}. The second column gives the abundance relative to the solar abundances recently determined using 3D calculations by \citet{ags2005}. The relative $1 \sigma$ errors are given in column 3.}
\begin{center}
\begin{tabular}{r|ccc}
      & $M/M_{\odot 1D}$ & $M/M_{\odot 3D}$ & $1 \sigma$ (dex) \\\hline
C   &  0.32  &  0.43  &  0.10 \\
N   &  0.98  &  1.35  &  0.13 \\
O   &  0.44  &  0.65  &  0.14 \\
Mg &  0.54  &  0.60  &  0.21 \\
Al   &  0.36  &  0.45  &  0.16 \\
Si   &  0.36  &  0.40  &  0.23 \\
S    &  0.65  &  1.00  &  0.37 \\
Fe  &  0.55  &  0.62  &  0.23 \\
\end{tabular}
\end{center}
\label{tab:abun}
\end{table}%

$\beta$\,Cep is the primary star of a triple system. The visual companion ADS\,15032\,B, is a $V=7.9$ A2.5V star located at 13.4 arcsec from the primary. It has an orbital period of about 91.6 yr and was at periastron in 2006 (\citealp{pb92}). The third component of the system, a physical companion, is a $V=6.6$ B6-8 star situated at 0.1 arcsec from the primary (\citealp{sho+2006}), detected through speckle interferometry (\citealp{gls72}). 

\begin{figure}[!htbp]
\begin{center}
\resizebox{\hsize}{!}{\includegraphics[clip]{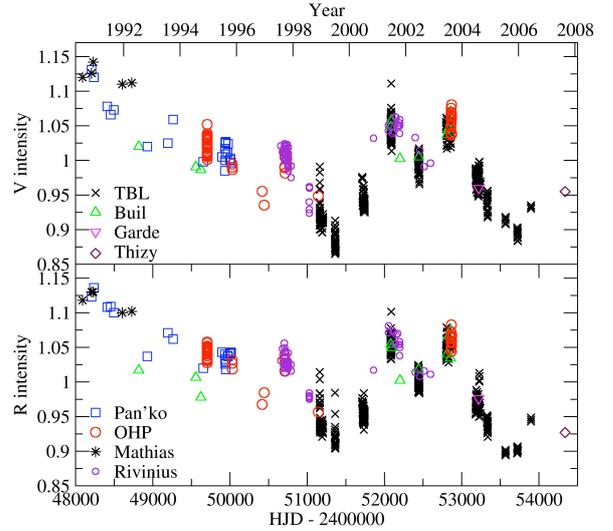}}\vspace{-4.5cm}
\caption{Evolution of the intensity of the Halpha emission of beta Cep with time. Data have been obtained at Telescope Bernard Lyot (TBL) of Pic du Midi Observatory and Observatoire de Haute-Provence (OHP), and collected from the litterature (\citealp{pt97}), amateurs astronomers (C. Buil, O. Garde, T. Thizy) and private communications (P. Mathias, T. Rivinius).}
\label{fig:Halpha}
\end{center}
\end{figure}

The $\beta$\,Cep system also exhibits emission in its Balmer lines, so that the primary itself has long been considered as a Be star. In Be stars the Balmer emission is due to the presence of circumstellar matter ejected by the star. The Be character is usually variable with time, and the \bcep\ system is no exception. The time evolution of the intensity of its H$\alpha$ emission line  is shown in Fig.~\ref{fig:Halpha}: major outbursts occured around 1990 and 2001. At the time of our X-ray observations in 2005 (see Fig.~\ref{fig:Halpha}), the H$\alpha$ emission was decreasing but still visible in the line profile. However, Be stars usually rotates fast, at about $v \sin i = 250$ km s$^{-1}$ on average, while the $\beta$\,Cep primary rotates intrinsically slowly, at $v \sin i$ = 20 km s$^{-1}$, and $i = 60$ deg (\citealp{alg2002}), making it different from classical Be stars. 

The condundrum relative to the unusually low rotation speed of \bcep\ relative to its Be status has apparently been recently solved by \citet{sho+2006}, who argue that the H$\alpha$ emission does not come from the bright primary, but rather from the fainter speckle companion. This would make the \bcep\ primary a normal B star, rather than a peculiar Be one. 

The rotational period of \bcep\ is almost exactly 12 days, and in the present study we use the ephemeris of D01,

\begin{equation}
JD = 2\,451\,238.15 + N \times 12.00092
\end{equation}
where phase 0 is the phase at which the longitudinal magnetic field (see below) is at maximum. This ephemeris is consistent, within the error bars, with the values published by \citet{hjd+2000} and \citet{hst2005}. 

\subsection{The magnetic field of $\beta$\,Cep}

Over the past decade magnetic fields have been detected in an increasing number of normal massive stars, in addition to the well known fields of Ap/Bp stars. In particular magnetic fields have been discovered in a few $\beta$\,Cep stars including \bcep\ itself (\citealp{hjd+2000}). The longitudinal magnetic field of \bcep\ is observed to vary with the 12 day rotation period.
 
\begin{figure}[!htbp]
\begin{center}
\resizebox{\hsize}{!}{\includegraphics[width=8cm]{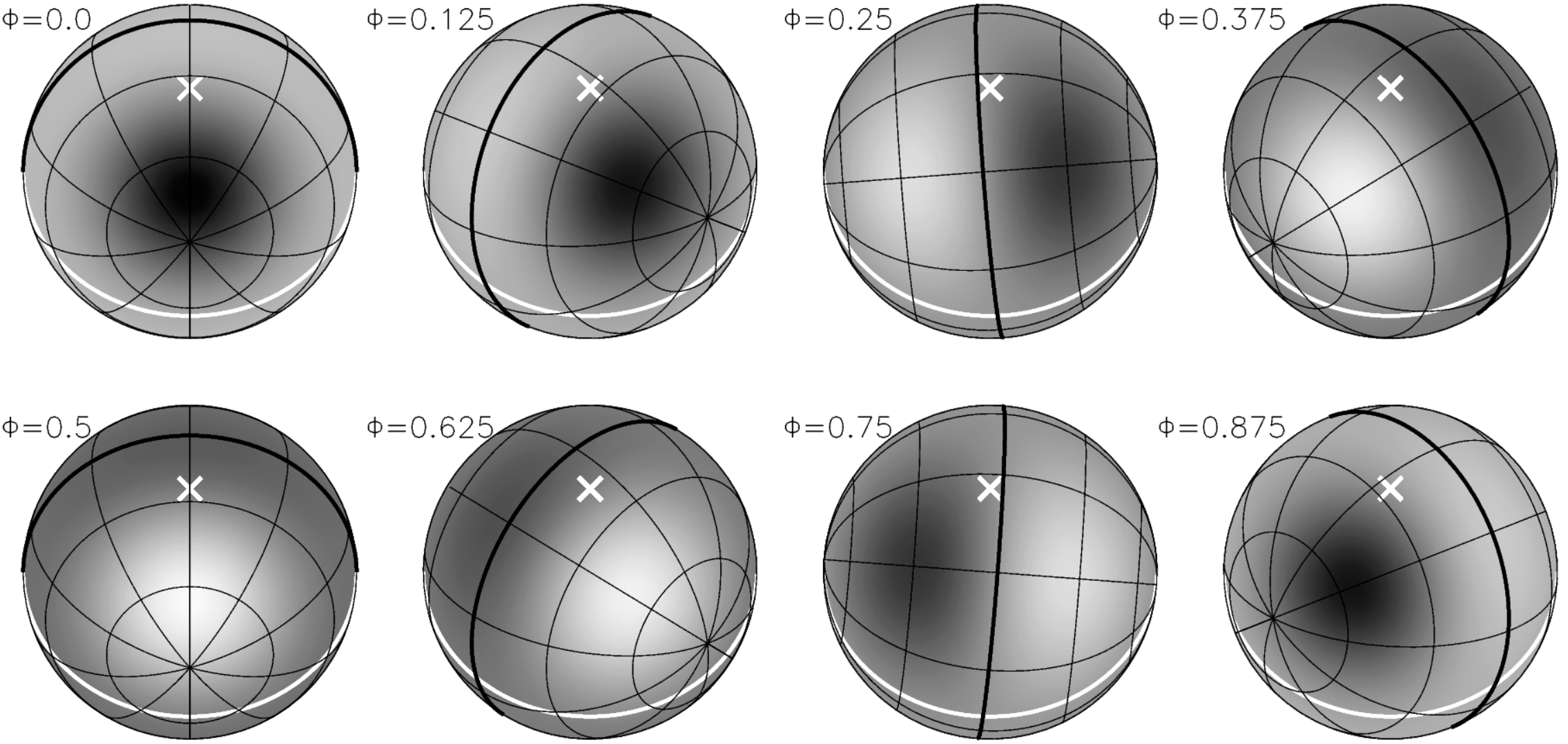}}
\caption{Greyscale representation of the relative contribution of the magnetic
  dipole to the integrated longitudinal magnetic field on the visible hemisphere
  of $\beta$\,Cep, at different rotational phases, with step $\delta\Phi=0.125$.
  Phase 0 is the phase at which the longitudinal field is at maximum, as in the
  ephemeris above. The black color corresponds to positive field values and the
  white color to negative field values. A grid of magnetic longitudes and
  latitudes is overplotted, with the magnetic equator shown as a thicker black
  line. The rotation axis and equator are shown with a white cross and line
  respectively. Although the strongest magnetic field is at the magnetic poles,
  the positions on the stellar surface that contribute the most to the
  longitudinal field are not at the poles, due to a geometrical effect and to
  the limb darkening effect.}
\label{dipole_model}
\end{center}
\end{figure}

These observations of magnetic massive stars can usually be explained with the magnetic oblique rotator model (\citealp{sti50}). In this model, the magnetic field structure is not symmetric about the rotation axis of the star. For the simplest case of a dipole field, this means that the axis of the dipole and the axis of rotation of the star do not coincide. The observed geometrical configuration can then be characterized by the inclination angle $i$ between the observer's line-of-sight and the stellar rotation axis, and by the obliquity angle $\beta$ between the magnetic axis and the rotation axis. In the case of a dipolar magnetic star, as the star rotates, the aspect of its visible hemisphere changes. This leads to variations in a number of observables with the stellar rotation period. $\beta$\,Cep hosts such a magnetic oblique dipole field with an inclination angle $i=60^{\circ}$ and its rotation and magnetic axis almost perpendicular ($\beta=85^{\circ}$). Therefore, as the star rotates, the observer successively sees the magnetic equator nearly edge-on or face-on. Fig.~\ref{dipole_model} shows the magnetic oblique dipole model of $\beta$\,Cep at different rotation phases, with the magnetic equator drawn as a thicker black line.

In the presence of a sufficiently strong dipolar magnetic field, the stellar wind escapes from the star via its magnetic poles and follows the magnetic field lines. As the star rotates one thus observes a variation of the equivalent width of the wind-sensitive lines such as the UV resonance lines of highly ionized species. Such variations have been observed  in $\beta$\,Cep, by \cite{hjd+2000}, who find that the C\,{\sc iv} line is strongly modulated with the rotation period of 12 d, maintaining phase coherence over several years. This shows that the transition-region temperature material responsible for the emission in C\,{\sc iv} is magnetically confined (and, incidentally, that it is associated with the primary, rather than with the Be secondary). 

If the magnetic field is strong enough the wind can be fully magnetically confined in a region close to the star (\citealp{bm97b}). In this case the stellar wind particles coming from both magnetic poles, forced to follow the magnetic field lines, collide at the magnetic equator creating a decretion disk or circumstellar clouds. The MCWS model has been successfully applied to explain the characteristics of the X-ray emission in a number of massive stars, in particular $\theta^1$ Ori C (\citealp{dbh+2002}; \citealp{goc+2005}). D01 showed that the MCWS seems to also apply for $\beta$\,Cep. It explains at the same time the rotation modulation observed in the wind-sensitive UV lines and in the longitudinal field measurements.

\subsection{The X-ray emission of $\beta$\,Cep}

As mentioned above, stars hotter than B4 have sufficiently strong winds to give rise
to X-ray emission in wind shocks. However, in the case of magnetically confined winds, shocks may also take place at the stellar equator, where the winds originating from the two magnetic poles collides. The resulting shock is expected to produce X-ray emission localized at or close to the magnetic equator. 

\bcep\ has been known to be an X-ray source since its observation with \emph{Einstein}. The X-ray spectrum, derived from ROSAT All Sky Survey observations obtained in 1990 and reported by \citet{bsc96}, has been interpreted as originating from a thermal source with an X-ray temperature of 0.24 keV and an X-ray luminosity of $6.5\times 10^{29}$ erg/s, for an assumed distance of 71 pc. Rescaling the reported X-ray luminosity to the Hipparcos distance of 182 pc, one obtains $L_{\rm X} = 4.2\times 10^{30}$ erg/sec.  \citet{asr+84} had obtained $L_{\rm X} = 9.1\times 10^{30}$ erg/sec, from data obtained with the IPC onboard Einstein, assuming a distance of 263 pc. Again rescaling this X-ray luminosity to the Hipparcos distance of 182 pc, one obtains $L_{\rm X} = 4.3\times 10^{30}$ erg/sec, which is very close to the value derived above from ROSAT data.

\cite{coh2000} claim the detection of a 7\% modulation in the X-ray emission of \bcep\ at the 4.57 hrs pulsation period, based on analysis of the ROSAT All Sky Survey data.

The MCWS \refcom{by \citet{bm97b}} makes clear predictions about the characteristics of the X-ray emission from $\beta$\,Cep, as extensively discussed in Sect. 4.2 of D01. The luminosity, temperature distribution and emission measure predicted by the plasma distribution of \refcom{this} model are, in the 0.1--2.0 keV range of the ROSAT PSPC, $7.2\times 10^{30}$ erg/sec, 3.2 MK and $4.5\times 10^{53}$ cm$^{-3}$ for the X-ray luminosity, temperature and emission measure respectively. These are in good agreement with the ROSAT observation (\citealp{bsc96}). 

Moreover, the \refcom{same model} makes an easily testable prediction regarding the temporal variability of the X-ray emission, for which strong modulation as a function of rotational phase is expected. The X-ray emission is produced on the shocks located on both sides of the cooling disk, whose density is sufficiently high to make it completely opaque to the X-rays emitted by the shock. As a consequence, when the system is viewed with the disk edge-on, X-rays will reach the observer from both shocks (on the magnetic north and south side of the disk), whereas when the disk is seen face-on, only one of the shocks will be visible in X-rays. The expected X-ray modulation amplitude in the MCWS \refcom{scenario of D01} is thus of the order of 50\%, although (as discussed by D01) the predicted shape of the X-ray light curve can be strongly affected by the disk's characteristics, for example its warp.

\refcom{The later, dynamical models of \cite{uo2002} and \cite{goc+2005}, when applied to \bcep, make a very different prediction regarding the modulation of its X-ray emission: in particular, the magnetic confinement parameter }

\begin{equation}
\eta_* = \frac{B^2_* (\pi/2) R^2_*}{ \dot{M} v_\infty}
\end{equation}
\refcom{using the appropriate values for \bcep\ ($B = 360$ G, $R_* = 6.5\, R_\odot$, $\dot{M} = 10^{-9}\, M_\odot\,\rm{yr}^{-1}$, $v_\infty = 800$-$1500\,$ km s$^{-1}$) would be high ($\eta_* \simeq 1000$), so that the \cite{goc+2005} model predicts that the magnetosphere would be rigid out to the Alfven radius ($R_{\rm A} \simeq 5\, R_*$) and the X-ray shocks would form just outside the Alfven radius but inside the Kepler radius $R_{\rm K} = 7\, R_*$ (M.~Gagne, private communication). This implies a small modulation of the X-ray activity, at the $\simeq 5\%$ level, with a minimum when the star is seen magnetic equator on, i.e.\ with the opposite phase as the modulation predicted by the D01 model.}

\section{Observations}
\label{sec:obs}

In the course of our campaign we have observed \bcep\ with both \emph{Chandra} and \xmm. In both cases the proposed observational strategy was to perform 4 observations, as close as possible to phases 0.0, 0.25, 0.50 and 0.75. Phases 0.0 and 0.50 correspond to the magnetic equator seen face-on, while phases 0.25 and 0.75 correspond to the edge-on configuration. In the MCWS scenario, these 4 observations should ensure maximum modulation of the X-ray emission.

The \xmm\ observatory allows simultaneous operation of the RGS high-resolution spectrograph and of the EPIC cameras. The goal of the \xmm\ observation was to detect modulation in the spectral parameters, namely the global spectral temperature and emission measure as observed in the EPIC low-resolution CCD spectra, and in the triplet ratios from the RGS spectra. The triplet ratios, as discussed below, are a diagnostic, for early-type stars, of the distance at which the emitting plasma is located from the star. 

The \emph{Chandra} observation was performed using the LETGS spectrograph, with the aim of detecting whether the lines would be broadened asymetrically (as e.g.\ expected in a unconfined, self-shocked wind), or symmetrically (as expected if significant rotational broadening is present), or whether modulation of the radial velocity with the star's rotational period is present (as expected e.g.\ in the presence of strongly localized plasma).

In practice, a number of constraints on the scheduling of space observatories made it difficult to schedule the observations exactly at the desired phases. The \emph{Chandra} observations were particularly critical because of issues with the spacecraft thermal control, so that in practice the phase coverage is, specially for the \emph{Chandra} data, not optimal -- see Table~\ref{tab:pnfits} for the phase coverage of the \xmm\ observations, and Table~\ref{tab:chandraobs} for the \emph{Chandra} observations.
 
 \subsection{\xmm\ data reduction}
    
Each of the 4 \xmm\ observations was processed individually with the standard SAS V.6.0 pipeline. We have reduced and analyzed the RGS (1 and 2) data, as well as the EPIC MOS and pn data. We will report only the results obtained on the pn data, as these were consistent with the MOS data (with one exception noted in the text). Some periods of high background (proton flaring) were present in the observations, so that the EPIC data were screened to exclude them.  We used as criterium the number of events detected at energies over 10 keV in the whole frame, and we discarded data periods in which this rate was greater than 12 per minute. The RGS1 and RGS2 spectra were analyzed separately, using the ``PintOfAle'' package (\citealp{kd2002}) to determine the line fluxes.

Given the accuracy of the pointing, we found that it was possible to use the same source and background regions for all four observations.

\begin{figure*}[htbp]
\begin{center}
%\vspace{-2.cm}\hspace {-1cm}
\epsfig{file=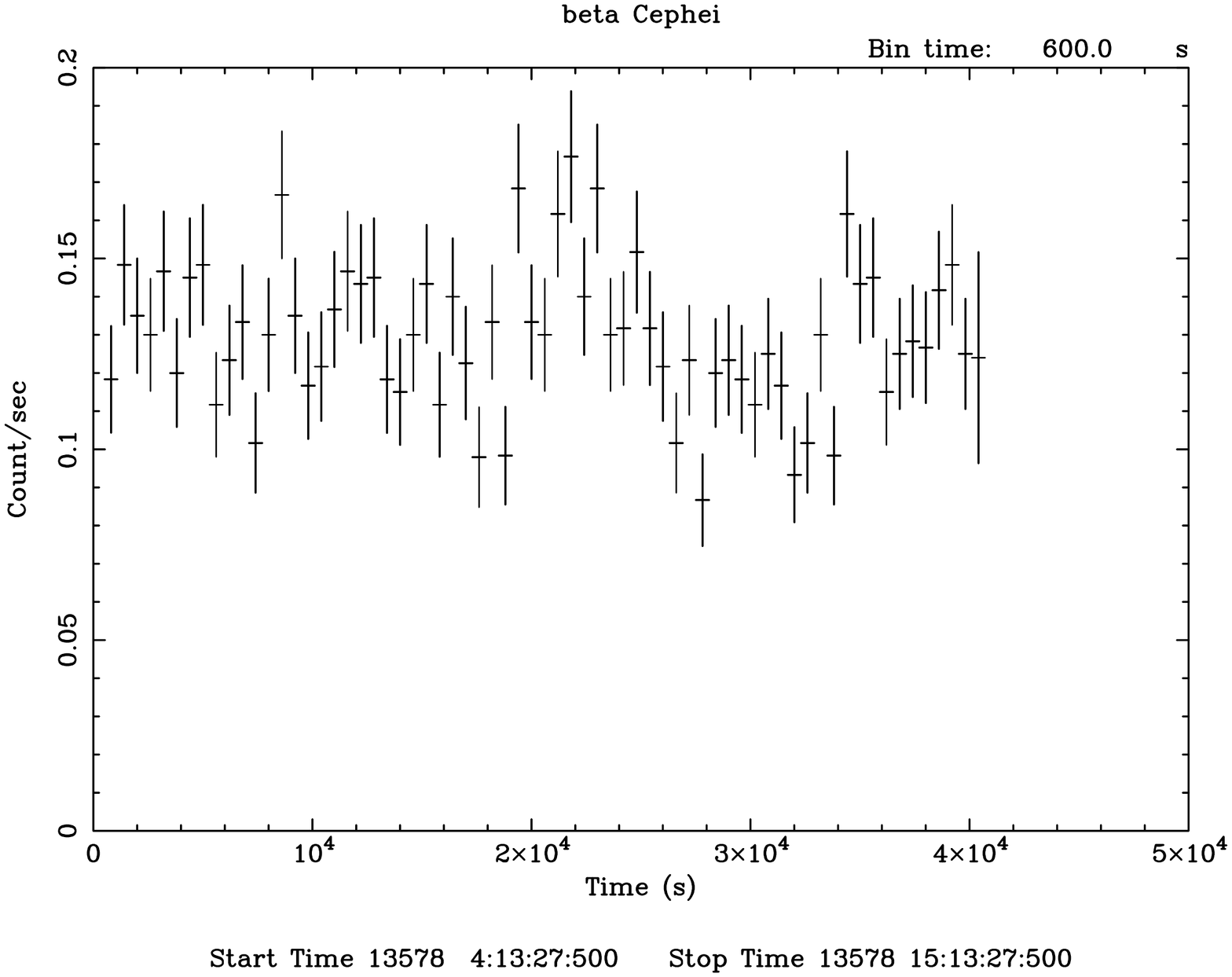, width=7cm, angle=-90}\epsfig{file=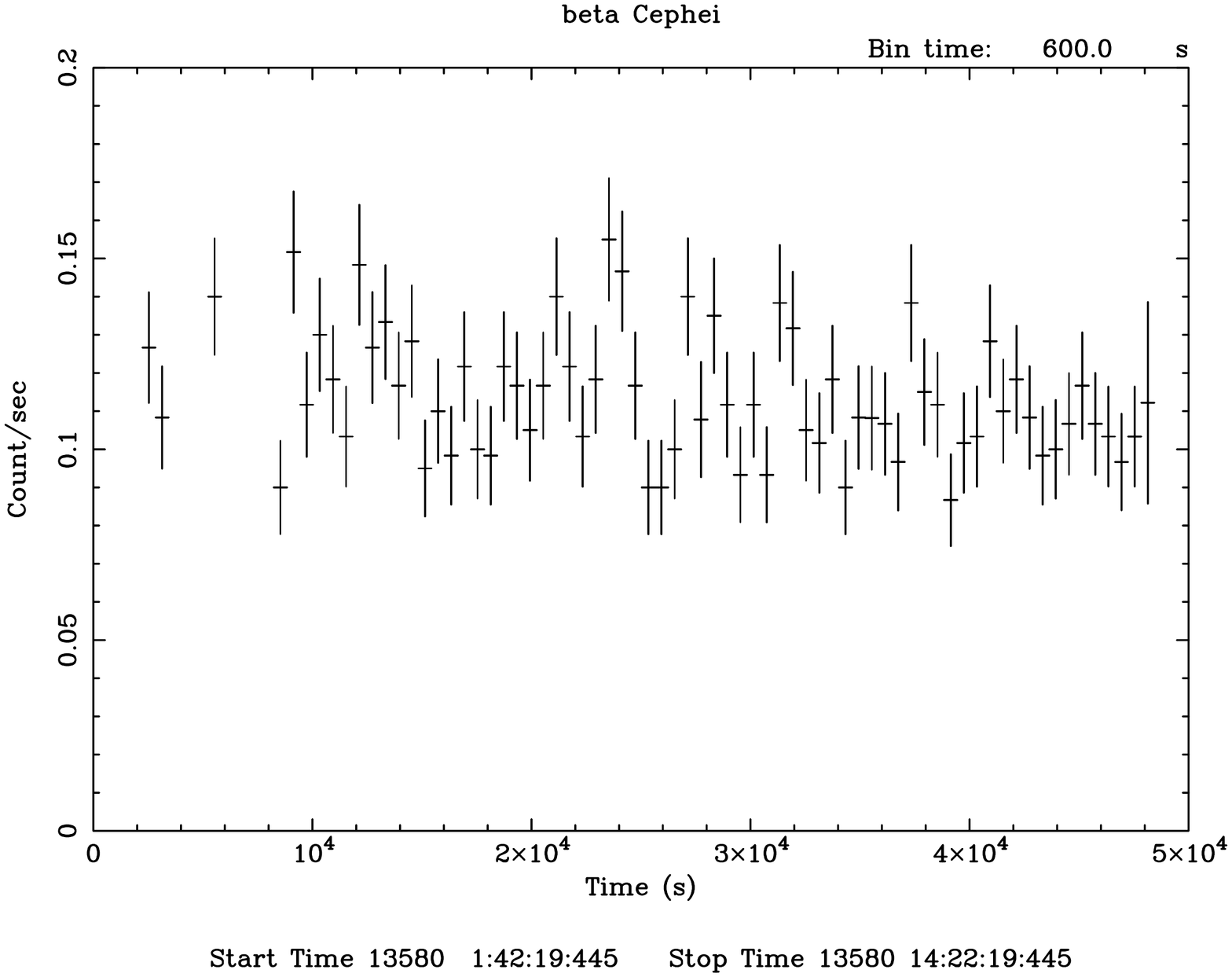, width=7cm, angle=-90}\\\vspace{-7mm}
\epsfig{file=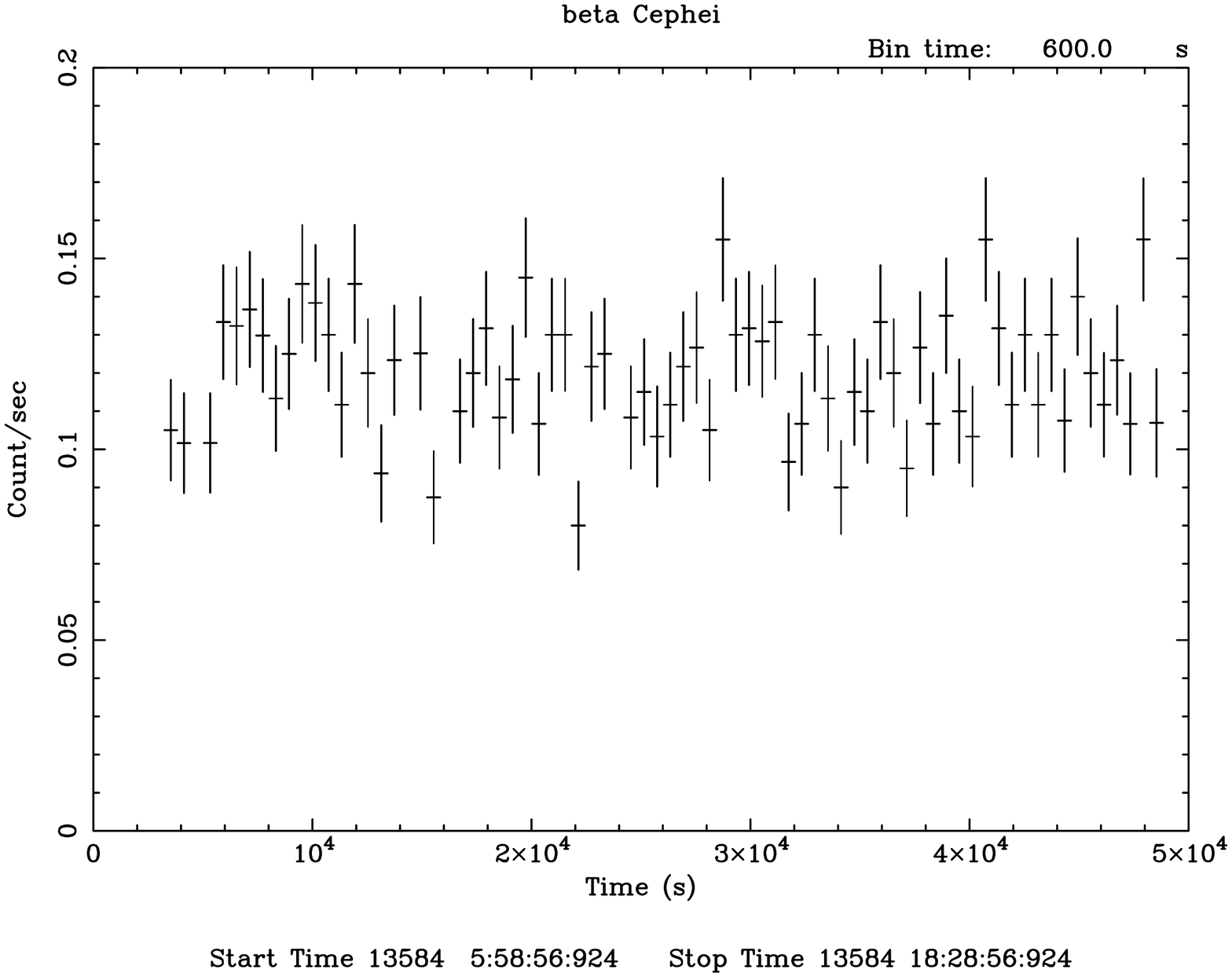, width=7cm, angle=-90}\epsfig{file=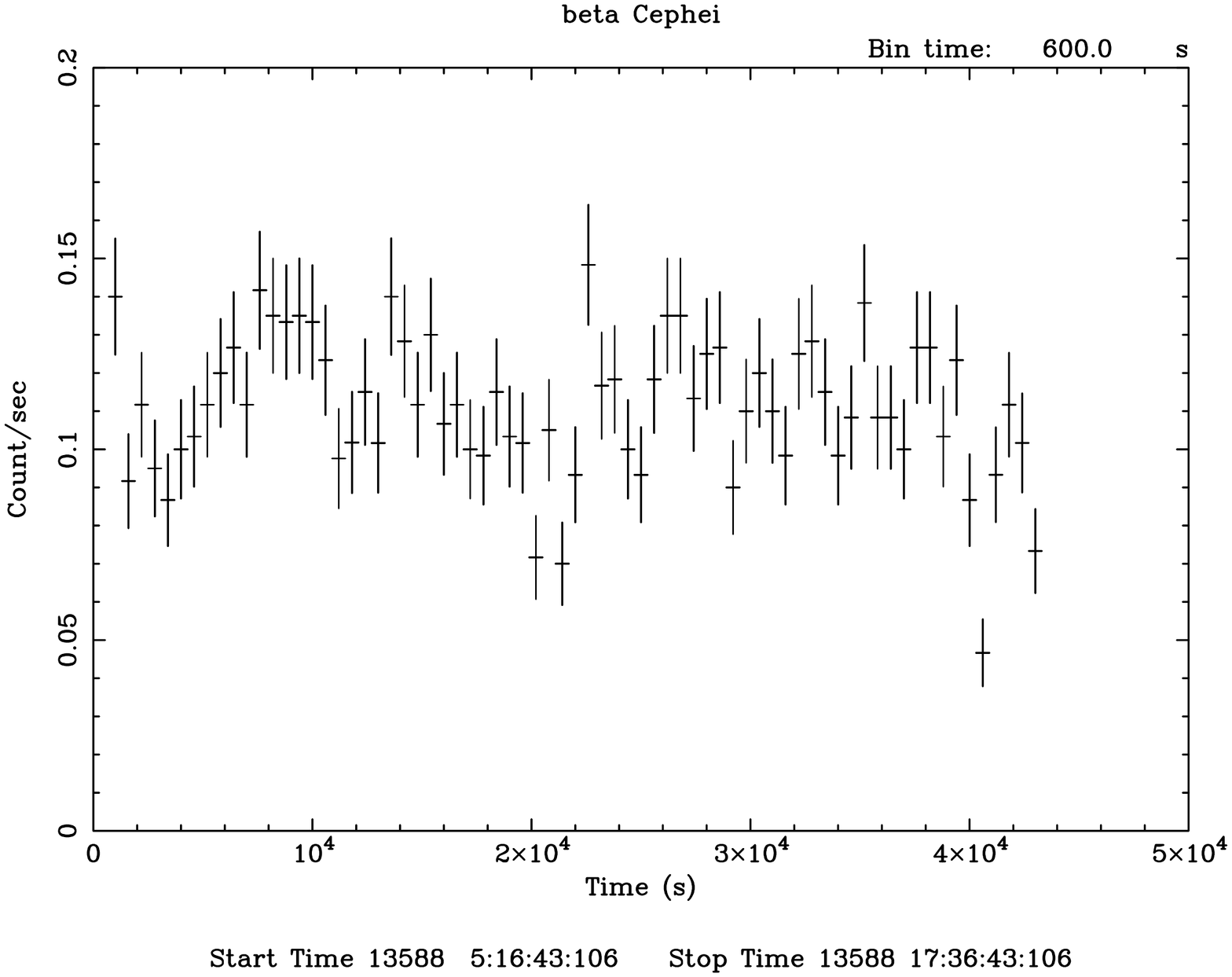, width=7cm, angle=-90}
%\vspace{-6.3cm}
\caption{The 4 light curves from the EPIC (pn) observations, background subtracted and binned at 1200 s resolution. }
\label{fig:pnlc}
\end{center}
\end{figure*}

\subsection{Chandra data reduction}

The {\em Chandra} data were reduced using the standard pipeline procedures in the latest release of the data reduction package, CIAO (v.3.3.0.1).  Barycentric corrections were applied to the timing in the reduced files, accounting and correcting for the spacecraft motion. Background-subtracted spectra and lightcurves were extracted from four separate event files, one for each {\em Chandra} observation. X-ray lightcurves were created by defining source and background regions and dividing the background region count rate by the relevant scaling factor prior to background subtraction. The pointing stability of \emph{Chandra} was also stable enough to enable the use of the same background and source regions for all four exposures. 

Four sets of spectra were extracted, one for each of the four observations,  integrating over the full exposure in each observation. As we wish to measure centroid positions accurately and individual lines have limited statistics, we sum the spectra from observations taken at similar phases (the first two data sets are centered near phase $\sim 0$ and the other 2 near phase 0.25). As the LETG wavelength scale is subject to non-linear deviations from the laboratory positions of the wavelengths (\citealp{cdk+2004}), the $+1$ and $-1$ orders may show systematic differences. To ensure that we can distinguish between instrumental variability and genuine variability in the target system, we consider the $+1$ and $-1$ orders separately in the subsequent analysis. We use the $+1$ and $-1$ order profiles to test for consistency and evaluate the precision of any shifts in centroid position measurements from one observation to another. 
  
\section{Results}
\label{sec:res}

\subsection{EPIC data}
  
\begin{figure*}[htbp]
\begin{center}
%\vspace{-2.cm}\hspace {-1cm}
\epsfig{file=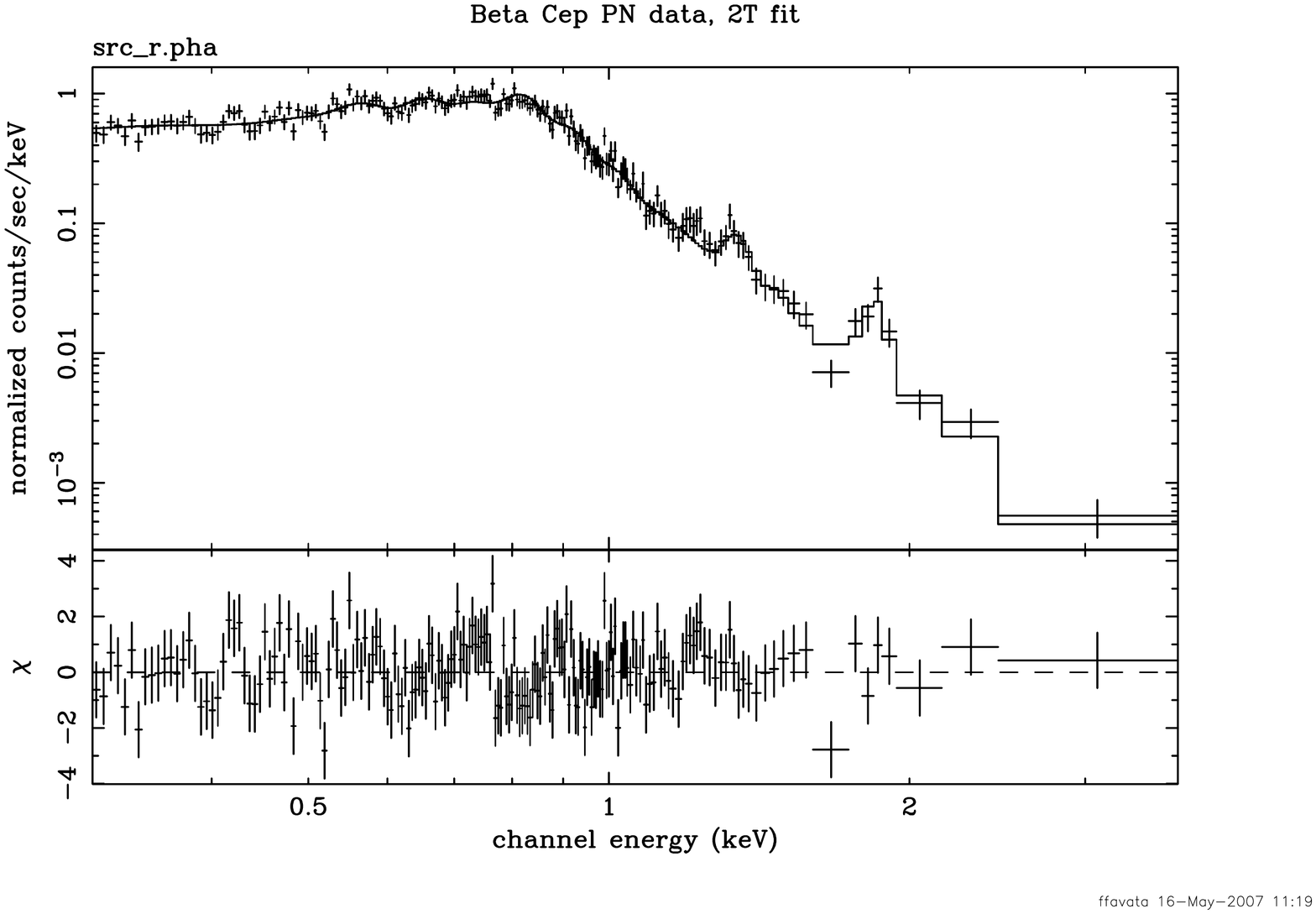, width=7cm, angle=-90}\epsfig{file=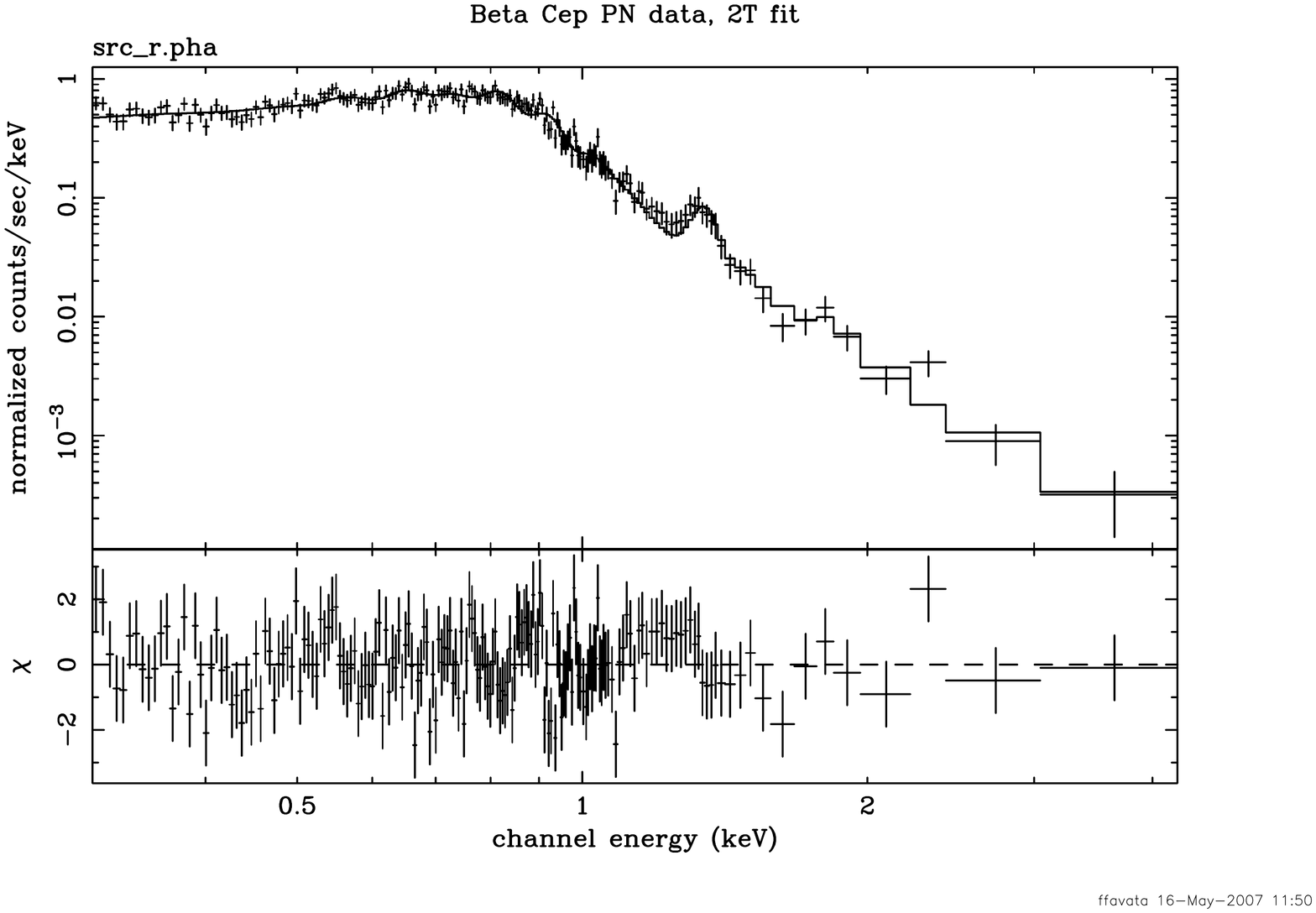, width=7cm, angle=-90}
%\vspace{-6.3cm}
\caption{Two-temperature, variable abundance fits to the EPIC pn spectra obtained in the first (left panel) and second (right panel) observations of \bcep. The relevant best fit parameters are presented in Table~\ref{tab:pnfits}. }
\label{fig:pnfits}
\end{center}
\end{figure*}

We first inspected the background-subtracted light curves from the pn detector, shown in Fig. \ref{fig:pnlc}. Small amplitude variability is likely present in all 4 observations, with a formal Kolmogorov-Smirnov (KS) test showing a high probability of the source having undergone some actual variability during the 4th segment (bottom right panel of Fig.~\ref{fig:pnlc}). On the other hand, during the 3rd segment of the observation (bottom left panel of Fig.~\ref{fig:pnlc}), the KS test gives a very low probability of the source being variable, consistent with a lack of actual variations during the observation. 

We also searched for evidence of modulation in the intensity of the X-ray emission at the 4h 34m pulsation period of \bcep, as reported by \cite{coh2000}. A period search performed on the complete pn data set using a $\chi^2$ test on the phase-folded data set failed to detect any significant periodicity. In particular, no periodicity is evident at the 4h 34m pulsation period. \refcom{Inspection of the 4 pn individual light-curves phase-folded at the pulsation period does not show any evidence of modulation in any of them. Furthermore, any possible hint of a modulation is at a different phase in each of the 4 light curves.} This, jointly with the fact that \cite{coh2000} have not performed a period search, but rather assumed the period and fitted the amplitude and phase, makes it likely that the period reported by \cite{coh2000} is spurious.

We have fitted the 4 individual EPIC pn spectra with variable abundance thermal models. To obtain a satisfactory fit the spectra require two temperature components. The lower temperature component at $T\simeq 0.25$ keV is dominant, with an emission measure in all cases more than one order of magnitude larger than the one of the hotter temperature. The hotter temperature is not well constrained, due to the low statistics of the high-energy tail in the spectra, and it ranges between 0.7 and 2.0 keV. Abundances are also not strongly constrained, and appear to be, if anything, moderately depleted with respect to the photospheric abundances of \bcep. The best fit values reported in Table \ref{tab:pnfits} show some evidence for variations of the abundance among the four spectra, in particular for Si. While the formal robustness of such fluctuations is modest, certainly the visual appearance of the pn spectra for the first and second observation (Fig.~\ref{fig:pnfits}) is strikingly different, with the Si complex at $\simeq 1.9$ keV very prominent in the first observation and almost absent in the second one.

The level of photometric variability among the 4 observations is modest (of order 10\%), allowing us to perform a joint spectral analysis of all four spectra simultaneously. The results are reported in Table~\ref{tab:pnspecall}. Given the availability of recently determined photospheric abundances for \bcep\ (\citealp{mba+2006}, see Table~\ref{tab:abun}) we can determine the ratio of coronal to photospheric metallicity for \bcep\ itself, without reference to the solar metallicity, as reported in Table~\ref{tab:abun}. Only 3 elements (O, Si, Fe) have both coronal and photospheric abundances available, and again, moderate metal depletion in the X-ray emitting plasma with respect to the stellar photosphere is present.

The $N({\rm H})$ resulting from the fit ($2.5\times 10^{20}$ cm$^{-2}$) corresponds (using the relation $N({\rm H})/A_V = 1.9 \times 10^{21}$) to an interstellar absorption of 0.13 mag. This is consistent with the published values for the interstellar reddening toward \bcep\ ($E_{B-V} \le 0.04$, corresponding to $A_V \le 0.13$).

In the ROSAT PSPC band (0.1--2.0 keV), the best-fitting model to the pn data has a flux corresponding to an X-ray luminosity of $4.0\times 10^{30}$ erg/sec, showing very little long-term variability when compared to the 1990 ROSAT observation, which had an X-ray luminosity of $4.2\times 10^{30}$ erg/sec. 

\begin{figure}[htbp]
\begin{center}
\vspace{-1.7cm}
\epsfig{file=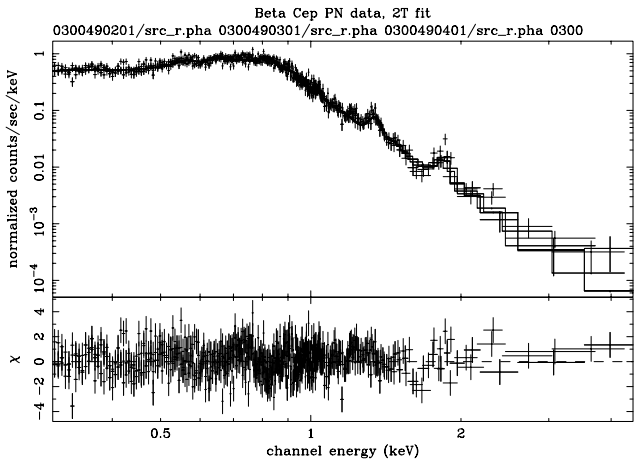,height=10cm, angle=-90}\vspace{-0.7cm}
\caption{Joint two-temperature, variable abundance fit to the 4 pn spectra of \bcep.The relevant best fit parameters are presented in Table~\protect{\ref{tab:pnspecall}}. }
\label{fig:pnsimfits}
\end{center}
\end{figure}

\begin{table*}[htdp]
\caption{The spectral parameters derived from the \xmm\ EPIC pn observations, together with the relative phase coverage. The reported flux is measured in a 0.3-7.0 keV band.}
\begin{center}
\begin{tabular}{lrrrr}
\hline
Obs. & 201 & 301 & 401 & 501 \\\hline
Start date & 2005-07-27 & 2005-07-29 & 2005-08-02 & 2005-08-06 \\
Start time (UT) & 04:07:27 & 01:36:19 & 05:53:00 & 05:10:42 \\
Obs. duration (s) & 39\,234 &  41\,030 & 42\,504 & 38\,170 \\
$\phi$ begin & 0.03 & 0.19 & 0.54 & 0.87 \\
$\phi$ end    & 0.07 & 0.23 & 0.58 & 0.91 \\
$N({\rm H})$ ($10^{20}$ cm$^{-2}$) & $4.5\pm0.1$ & $5.2\pm0.2$ & $3.2\pm0.1$ & $5.8\pm0.2$ \\
$T_1$	(keV) & $0.23\pm0.06$ 	& $0.24\pm0.01$	& $0.29\pm0.01$	& $0.24\pm0.01$\\
$T_2$	(keV) & $0.69\pm0.03$	& $1.14\pm0.1$	& $2.0\pm1.4$	& $0.63\pm0.02$\\
$EM_1$ ($10^{53}$ cm$^{-3}$)	& $12.0\pm5.0$	& $18.8\pm6.4$	& $16.0\pm4.2$ & $17.2\pm6.4$\\
$EM_2$ ($10^{53}$ cm$^{-3}$)	& $0.96\pm0.42$	& $0.45\pm0.12$	& $0.14\pm0.20$ & $0.83\pm0.42$\\
flux($10^{-12}$ erg cm$^{-2}$ s$^{-1}$)	& 1.01	& 0.89	& 0.99	& 0.92\\
O/O$_\odot$ & $0.13\pm0.03$	& $0.08\pm0.02$	& $0.08\pm0.02$	& $0.09\pm0.02$\\
Ne/Ne$_\odot$ & $0.18\pm0.05$	& $0.22\pm0.06$	& $0.16\pm0.03$	& $0.16\pm0.03$\\
Si/Si$_\odot$ & $0.72\pm0.24$	& $0.23\pm0.13$	& $0.40\pm0.13$	& $0.41\pm0.16$\\
Fe/Fe$_\odot$ & $0.45\pm0.13$	& $0.28\pm0.08$	& $0.17\pm0.03$	& $0.32\pm0.08$\\\hline
\end{tabular}
\end{center}
\label{tab:pnfits}
\end{table*}

\begin{table}[htdp]
\caption{The joint best-fit spectral parameter for the 2-$T$ fit with individually varying metal abundances for the 4 pn spectra. The notation ``M/M$_\odot$'' indicates the ratio between the \bcep\ coronal abundance and the solar photospheric abundance. The notation ``M/M$_{\rm phot}$ indicates the ratio between the coronal abundance of \bcep\ determined here and the photospheric abundance of \bcep\ determined by \citet{mba+2006}. In all cases the solar photospheric abundances of \citet{gs98} have been used. The indicated uncertainties are 90\% errors as resulting from a $\Delta \chi^2$ analysis. For all heavy elements not explicitly indicated, the abundances are equal to the value for Fe.}
\begin{center}
\begin{tabular}{c|l}
$N({\rm H})$ & $2.50 \pm 0.01 \times 10^{20}$ cm$^{-2}$\\
$T_1$ & $0.24\pm0.01 $ keV\\
$T_2$ & $0.63\pm0.03 $ keV\\
$E\!M_1$ & $1.1\pm0.2$ $10^{54}$ cm$^{-3}$ \\
$E\!M_2$ & $1.3\pm0.3$ $10^{53}$ cm$^{-3}$ \\\hline
O/O$_\odot$ & $0.12\pm0.02 $\\
Ne/Ne$_\odot$ & $0.18\pm0.03$ \\
Si/Si$_\odot$ & $0.31\pm0.06 $\\
Fe/Fe$_\odot$ & $0.33\pm0.04$\\\hline
O/O$_{\rm phot}$ & 0.28\\
Si/Si$_{\rm phot}$ & 0.86\\
Fe/Fe$_{\rm phot}$ & 0.60\\
\end{tabular}
\end{center}
\label{tab:pnspecall}
\end{table}

\subsection{RGS data}	
\label{RGS data}

\begin{figure}[htbp]
\begin{center}
\vspace{-1cm}\hspace{-1.5cm}\epsfig{file=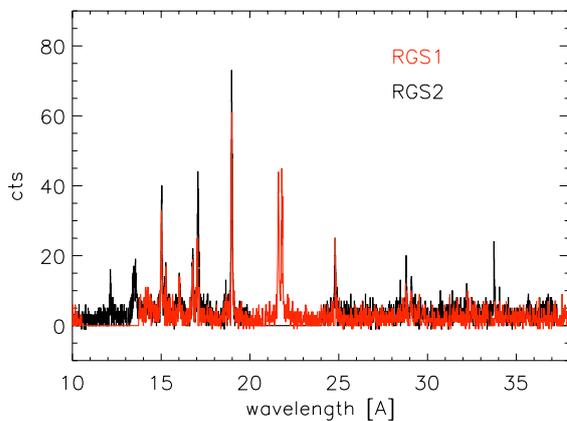, width=10.5cm}\vspace{-6.5cm}
\caption{The RGS1 (red) and RGS2 (black) spectra. All 4 individual spectra have been coadded in this plot. }
\label{default}
\end{center}
\end{figure}

\begin{figure}[htbp]
\begin{center}
\vspace{-1cm}\epsfig{file=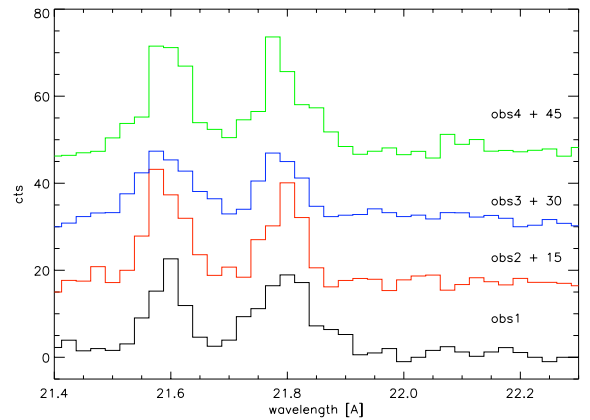, width=10.cm}\vspace{-6.cm}
\caption{The 4 individual RGS1$+$RGS2 spectra in the region of the O\,{\sc vii} triplet. The spectra have been offset in the vertical direction for clarity.}
\label{fig:orgs}
\end{center}
\end{figure}

\begin{figure}[htbp]
\begin{center}
\vspace{-6cm}\hspace{-1.cm}\epsfig{file=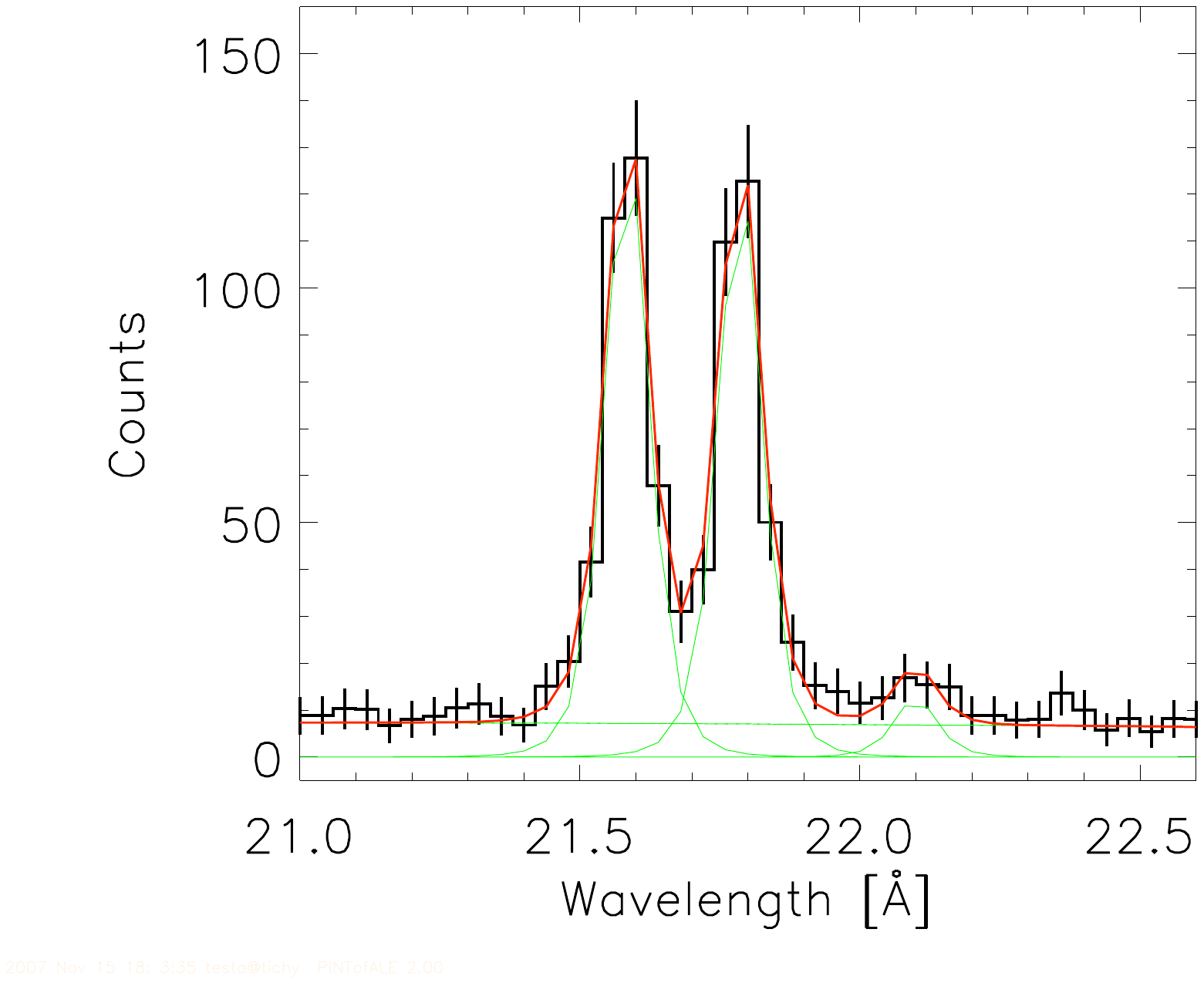, width=10.cm}\vspace{-0.5cm}
\caption{The 4 coadded RGS1$+$RGS2 spectra in the region of the O\,{\sc vii} triplet, with the fit to the 3 lines \refcom{plus a constant}.}
\label{fig:ocoadd}
\end{center}
\end{figure}

We have first analyzed the 4 individual RGS spectra separately. As shown in Fig.~\ref{fig:orgs} for the region of the \ovii\ triplet, given the moderate statistics of the spectra no difference is visible among the different spectra. We have therefore coadded them and limited the further analysis to the coadded spectrum. The coadded spectrum is shown in Fig.~\ref{fig:ocoadd}, again limited to the \ovii\ triplet region.

The RGS spectral range includes a number of He-like triplets which provide useful diagnostics of the plasma conditions. The most relevant at the plasma temperature of \bcep\ are the ones from \nvi\ (at 29 \AA), \ovii\ (at 22 \AA) and \neix\ (at 13.5 \AA), each composed of 3 spectral features: the resonance line $r$, the intercombination doublet $i$ (which is blended into a single observed line) and the forbidden line $f$. The intensity ratio $f/i$ is determined by both the plasma density (above a certain critical density which changes from element to element) and on the ambient UV field. The $f$ line is the product of the decay from a metastable state; at low density and in the absence of an ambient radiation field the metastable state will decay undisturbed and the $f$ line will be at maximum strength. Both collisions and radiative pumping can depopulate the metastable state, so that an increase in density or in the ambient radiation field will result in a less intense $f$ line. For the triplets of interest, the radiative pumping depopulating the metastable state is due to ambient UV photons. While in cool stars the $f/i$ ratio is commonly used as diagnostic of plasma density, in hot stars (such as \bcep) the ambient UV field due to the star's photospheric emission is strong and dominates the observed $f/i$ ratio for the typical plasma densities. The $f/i$ ratio depends on density and UV field through the relation

\begin{equation}
R = R_0 \frac{1}{1+\phi/\phi_{\rm c} + n_e/n_{\rm c}}
\end{equation}

where $R_0$ is the $f/i$ ratio in the limit of low-density and negligible UV field, $n_e$ is the electron density of the plasma, $n_{\rm c}$ is the critical density  (approximately, for N\,{\sc vi}, $5\times 10^9$, for O\,{\sc vii} $3\times 10^{10}$ and for Ne\,{\sc ix} $6 \times 10^{11}$ cm$^{-3}$), $\phi$ is the UV photoexcitation rate of the relevant transition (which depends on the ambient UV radiation field due to the star's photospheric emission, at the wavelength of 1906, 1637 and 1270 \AA\ respectively for \nvi, \ovii\ and \neix), and $\phi_{\rm c}$ is the rate at which $R = R_0/2$. To determine the photoexcitation rate $\phi$ we have kept into account the variation of the UV flux with distance from the photosphere, using the formalism of \citet{ms78} who embed this effect in the so-called dilution factor.

Under the assumption that the plasma is below the critical density $n_{\rm c}$, the observed $f/i$ ratio depends only on the distance from the photosphere, and therefore the observed value can be used to indicate the characteristic distance at which the bulk of the emitting plasma is located. Given that the expected density from e.g.\ the MCWS model are below $10^9$ cm$^{-3}$ we expect this assumption to be verified.

\begin{table}[htdp]
\caption{The line fluxes measured in the RGS spectrum. Fluxes are in $10^{-6}$ photons cm$^{-2}$ s$^{-1}$.}
\begin{center}
\begin{tabular}{rrrrr}
Det. & Line ID & $\lambda$ & flux & error \\\hline
RGS2   & Ne\,{\sc x} Ly$\alpha$          &  12.132   &  10.6  &   1.7 \\
RGS2   & Ne\,{\sc ix} $r$                &  13.447   &  10.4  &   1.7 \\
RGS2   & Ne\,{\sc ix} $i$                &  13.550   &  8.8   &   1.7 \\
RGS2   & Ne\,{\sc ix} $f$                &  13.700   &  1.6   &   0.9 \\
RGS2   & Fe\,{\sc xvii}                  &  15.010   &  21.7  &   2.2 \\
RGS2   & Fe\,{\sc xvii}                  &  15.190   &  6.1   &   1.4 \\
RGS2   & Fe\,{\sc xvii}                  &  15.260   &  8.9   &   1.6 \\
RGS2   & Fe\,{\sc xviii}+O\,{\sc viii} Ly$\beta$ &  16.010   &  6.6   &   1.9 \\
RGS2   & Fe\,{\sc xvii}                  &  16.070   &  3.6   &   1.9 \\
RGS2   & Fe\,{\sc xvii}                  &  16.780   &  17.7  &   2.0 \\
RGS2   & Fe\,{\sc xvii}                  &  17.050   &  20.8  &   3.1 \\
RGS2   & Fe\,{\sc xvii}                  &  17.096   &  18.9  &   3.0  \\
RGS2   & O\,{\sc vii}                    &  18.627   &  6.8   &   1.5  \\ 
RGS2   & O\,{\sc viii} Ly$\alpha$        &  18.970   &  59.4  &   3.2\\ 
RGS1   & O\,{\sc vii} $r$                &  21.600   &  45.1  &   3.4 \\
RGS1   & O\,{\sc vii} $i$                &  21.800   &  41.3  &   3.2 \\
RGS1   & O\,{\sc vii} $f$                &  22.100   &  4.8   &   1.9 \\
RGS2   & N\,{\sc vii} Ly$\alpha$         &  24.780   &  21.2  &   2.2 \\
RGS2   & N\,{\sc vi} $r$                 &  28.790   &  18.4  &   2.3 \\
RGS2   & N\,{\sc vi} $i$                 &  29.080   &  17.2  &   2.3 \\
RGS2   & N\,{\sc vi} $f$                 &  29.530   &  2.4   &   1.8 \\
\end{tabular}
\end{center}
\label{tab:rgs}
\end{table}%

We have performed the analysis of the 3 triplets visible in our spectra, namely \nvi, \ovii\ and \neix. The fluxes for all major lines used in our analysis are reported in Table~\ref{tab:rgs}. For the \nvi\ triplet no forbidden line is detected in the spectrum. The upper limit to its flux translates into an upper limit on the $f/i$ ratio and thus on the characteristic distance of the plasma from the photosphere.

The \neix\ triplet is significantly affected by blending, with the blending lines not resolved at \xmm\ resolution. While the forbidden line is relatively isolated, the intercombination line is strongly affected by blending from Fe lines, which would require very high $S/N$ and resolution to be properly resolved (e.g.\ \citealp{nbd+2003}). In our spectra the $f$ line is not detected, again giving only an upper limit to the characteristic distance from the star.

In the O\,{\sc vii} triplet the forbidden line is detected, at the $\simeq 3 \sigma$ level, in the coadded spectra (Fig. \ref{fig:ocoadd}), resulting in a determination of the ratio $f/i = 0.095 \pm 0.033$. 

To determine the influence of the photospheric UV flux on the X-ray triplets, a measure of the star's UV field is needed. The simplest approach is to use the star's effective temperature (26\,000 K) and, under the assumption of a black body emitter, compute the resulting UV flux at the wavelength of interest. A more empirical approach relies on the measurement of the UV flux at the wavelengths of interest based on e.g.\ IUE spectra. We have compared the results obtained using both approaches.

To determine the UV flux at the 3 wavelengths of interest we have retrieved from the public archive an IUE observation of \bcep, from which the flux at Earth has been determined. To derive the radiation intensity at the stellar surface, the flux needs correcting for the effects of interstellar extinction. Extinction toward \bcep\ is modest, and the values in the literature range from $E_{B-V} = 0.04$ (\citealp{kt80}) to $E_{B-V} = 0.01$ (\citealp{ks93}). To assess the sensitivity of our analysis to the value of interstellar extinction we have considered both the case  $E_{B-V} = 0.04$ and a case for no measurable extinction. Using the UV interstellar extinction law reported by \cite{zom90} we have determined the extinction at each of the 3 wavelengths, determined the radiation intensity and derived the equivalent blackbody temperature.

\begin{table}[htdp]
\caption{The unabsorbed UV flux at Earth and intensity at the stellar photosphere, at the wavelength relevant for the radiative pumping of the $f$ transition for the \nvi, \ovii\ and \neix\ triplets, derived from an IUE spectrum of \bcep. $F_\lambda$ in units of erg cm$^{-2}$ s$^{-1}$ \AA$^{-1}$, $I_\lambda$ in units of erg cm$^{-2}$ s$^{-1}$ \AA$^{-1}$ sr$^{-1}$}
\begin{center}
\begin{tabular}{l|c|c|c|c}
&                      $E_{B-V}$ & \nvi\ (1906 \AA) & \ovii\ (1637 \AA) & \neix\ (1270 \AA) \\\hline
$F_\lambda$ &   0.00 & $3.8\times10^{-9}$ & $5.6\times10^{-9}$ & $1.1\times10^{-8}$ \\
$I_\lambda$ &   0.00 & $1.1\times10^{9}$ & $1.6\times10^{9}$ & $3.2\times10^{9}$\\
T  & 0.00 & 19\,995 & 21\,227 & 23\,969 \\\hline
$F_\lambda$ &   0.04 & $5.1\times10^{-9}$ & $7.4\times10^{-9}$ & $1.6\times10^{-8}$ \\
$I_\lambda$ &   0.04 & $1.5\times10^{9}$ & $2.1\times10^{9}$ & $4.6\times10^{9}$\\
T  & 0.04 & 21\,674 & 22\,701 & 25\,899 \\
\end{tabular}
\end{center}
\label{tab:uvflux}
\end{table}%

\begin{figure}[htbp]
\begin{center}
\vspace{-1cm}\epsfig{file=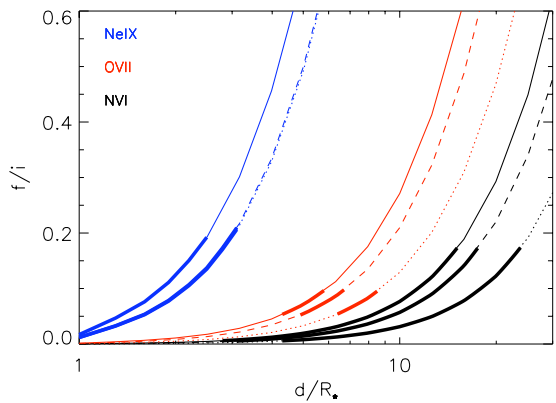,width=10.cm}\vspace{-6.cm}
\caption{The thick lines show the measured $1\sigma$ range for the $f/i$ line ratios, and indicate the allowed distance for the location of the X-ray emitting plasma for the \neix, O\,{\sc vii} and N\,{\sc vi} triplets. The continuous curve corresponds to the UV flux determined from the IUE spectrum with  $E_{B-V}=0.00$, the dashed curve to the same with $E_{B-V}=0.04$, and the dotted curve to the UV flux corresponding to a 26\,000 K blackbody. }
\label{fig:dist}
\end{center}
\end{figure}

The dependence of the $f/i$ ratio on the distance from the star's photosphere is plotted in Fig. \ref{fig:dist}, for the three different assumptions regarding the photospheric UV flux (black body, IUE spectrum with $E_{B-V}=0.00$, IUE spectrum with $E_{B-V}=0.04$). For each triplet the range of $f/i$ determined from the observed spectrum is indicated by a thicker line. For \nvi\ and \neix\ an upper limit to the location of the plasma can be determined, of respectively $\simeq 15$--$25\,R_{\star}$ for \nvi\ and $\simeq 2.5$--$3\,R_{\star}$ for \neix, depending on the assumption for the UV field. In the following we adopt the curve derived using the UV flux determined from the IUE spectrum and $E_{B-V}=0.04$. 

For \ovii\, the characteristic distance of the emitting plasma from the photosphere is between $\simeq 3$ and $\simeq 5\,R_{\star}$ using the UV flux determined from the IUE spectrum and $E_{B-V}=0.04$. If the \ovii\ emitting plasma is located at an effective distance from the photosphere of $\simeq 5\,R_{\star}$, the rotational velocity of the plasma, assuming rigid rotation at the star's rotational period of 12 d, would be 190 km/s, decreasing to 130 km/s for an effective distance of $\simeq 4\,R_{\star}$. Such velocity is below the velocity resolution of the \emph{Chandra} LETG/HRC instrument, and therefore one expects (as indeed observed) that the spectral lines in the LETG spectrum will be consistent with the instrumental profile.

The RGS spectrum also allows to easily derive abundance ratios by taking the ratios of lines of different elements characterized by similar emissivities as a function of temperature (\citealp{dt2005}). Using the Ne\,{\sc x} Ly$\alpha$ at 12.132\,\AA, Ne\,{\sc ix} $r$ line at 13.447\,\AA\ and the O{\,{\sc viii} Ly$\alpha$  line at 18.970\,\AA\ we have derived a ratio Ne/O of $0.32\pm0.05$ (about twice the solar ratio of 0.18 e.g.\ from \citealp{gs98}), marginally consistent within the error bars with the ratio derived from the pn spectrum ($\simeq 0.67$).  

\subsection{\emph{Chandra}/LETG observations}

The main aim of the \emph{Chandra} LETG observation of \bcep\ was to study the X-ray line profiles at the highest possible spectral resolution for the low-temperature plasma of \bcep, with the aim of detecting either a modulation in the line centroid or a broadening in the line profile. In the MCWS scenario modulation in the line centroid in phase with the rotation period would be expected if the plasma is distributed non-homogeneously around the star and at sufficient distance from the star so that the rotational velocity is visible as Doppler shift.

If on the other hand the plasma is distributed uniformly, no velocity shifts are expected, but stellar rotation will induce a broadened line profile. Assuming for \bcep\ $R = 6.5\,R_\odot$, and the rotational period of 12 d, the velocity of a plasma orbiting around the star will be $v = 28 \times r$ km sec$^{-1}$, where $r$ is the distance of the plasma from the stellar center expressed in stellar radii.

For this purpose we obtained four {\em Chandra} observations of $\beta$ Cep (ObsIDs 5395, 6138, 7194 and 7195) using the low energy transmission grating (LETG) with the high-resolution camera spectroscopic array (HRC-S). The phases spanned by the four {\em Chandra} exposures are listed in Table\,\ref{tab:chandraobs}, computed using the ephemeris of D01. While the original plan was to obtain 4 observations at phases as close as possible to 0.0, 0.25, 0.5 and 0.75, the current restrictions on \emph{Chandra} operations limited the accessible phases. As Table~\ref{tab:chandraobs} shows,  two datasets were taken near phase 0.0 and the final two observations were acquired near phase 0.25, providing a more limited phase sampling than originally planned, but still allowing to sample the system in both face-on and edge-on configurations. The resulting LETG spectrum, obtained by summing all the four observations together, is shown in Fig.~\ref{fig:letgspec}.

\begin{table*} 
 \centering 
% \begin{minipage}{90mm} 
%  \begin{tabular}{@{}lllllll@{}} 
  \begin{tabular}{lllllll} 

\hline 
ObsID & Exposure &  $T_{\rm start}$ & $T_{\rm end}$ & $\phi_{\rm start}$ & $\phi_{\rm end}$ & Count rate\\ 
 & ksec & JD & JD & & & cts/sec \\
\hline 
5395  & 36.43           & 2\,453\,674.4519389  & 2\,453\,674.8897456 & 0.0096      & 0.0461 & 0.086 \\ 
6138  & 29.54 & 2\,453\,675.0851294  & 2\,453\,675.4490259 & 0.0624      & 0.0927 & 0.089\\ 
7194  & 38.17 & 2\,453\,677.1208892  & 2\,453\,677.5798238 & 0.2320      & 0.2702 & 0.082 \\ 
7195  & 27.15 & 2\,453\,677.7494105  & 2\,453\,678.0855105 & 0.2844      & 0.3124 & 0.075\\ 
\hline 
\end{tabular} 
\caption{The observation log of the \emph{Chandra} data sets.}
%\end{minipage} 
\label{tab:chandraobs} 
\end{table*} 

\begin{figure}[htbp]
\begin{center}
\vspace{-2.cm}\hspace {-1cm}\epsfig{file=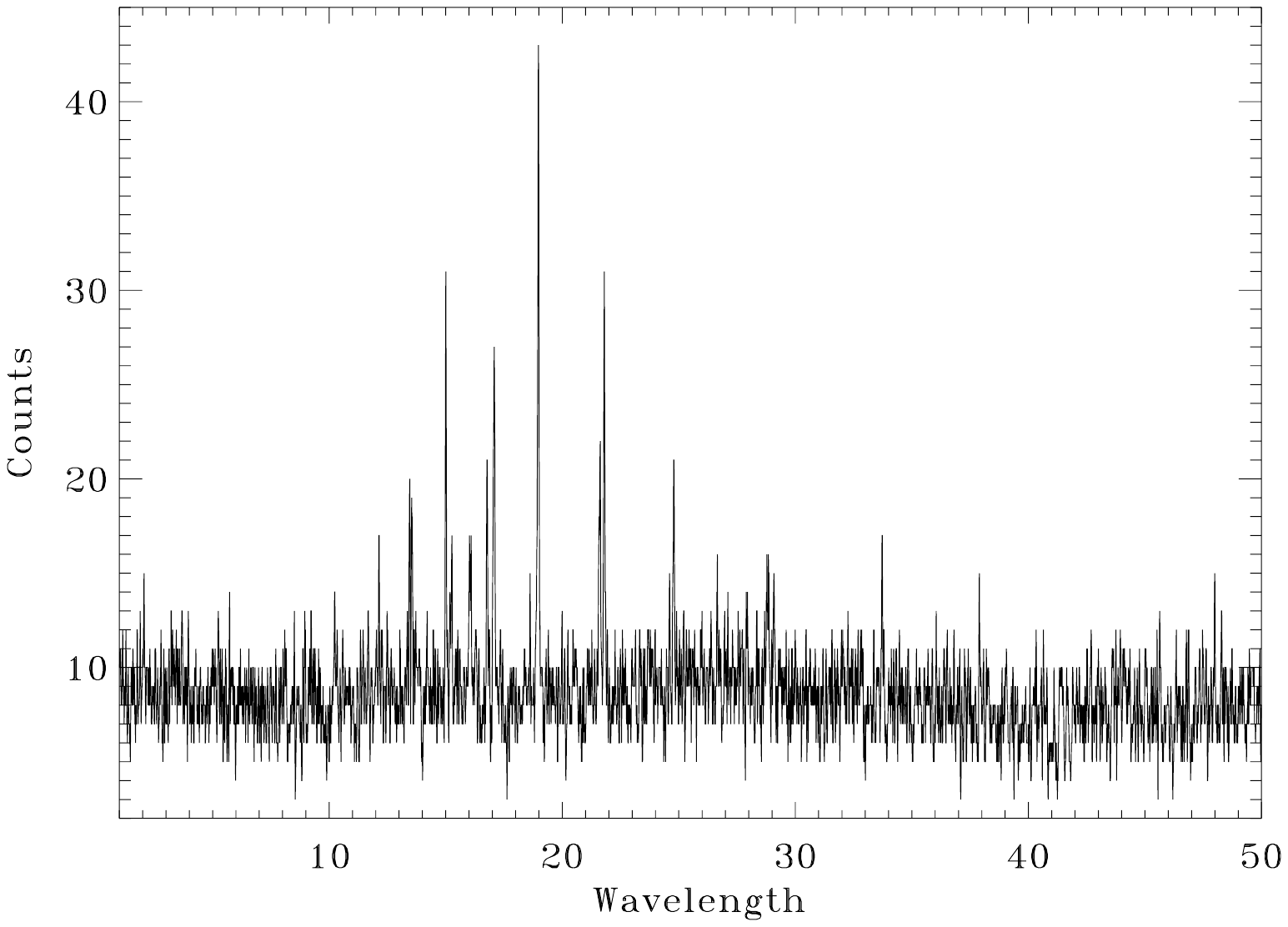, width=10cm, angle=0}\vspace{-6.3cm}
\caption{The LETG spectrum of \bcep\ obtained by summing together the four \emph{Chandra} observations discussed here. }
\label{fig:letgspec}
\end{center}
\end{figure}

The LETG/HRC-S configuration enables precise measurement of velocity shifts in X-ray spectra as the instrumental FWHM is 0.057 \AA, which corresponds to a velocity FHWM of approximately 900 km/s beyond a wavelength of 19 \AA. As shown e.g.\ by \cite{hbm2004}, the precision with which the centroid of a line profile can be measured depends strongly on the peak number of counts  in the line profile. By summing up the counts obtained over two exposures centered around similar phases, we can maximise our centroiding ability. Given the count rate of $\beta$ Cep, and the exposure times, our measurement error on the centroid position of the strongest 
unblended line profile, O\,{\sc viii} 18.97\AA, for observations 1 and 2 (centered at phases 0.05 and 0.27 respectively) corresponds to $\sim 160$ km/s. 

The spatial resolution of \emph{Chandra} can easily resolve \bcep\ from its companion at 13.4 arcsec. Visual inspection of the \emph{Chandra} image shows that no X-ray emission is visible from the companion, and that the X-ray source is coincident with the optical position of \bcep\ within a fraction of arcsec. The speckle companion at 0.1 arcsec is on the other hand not resolved.

We have first extracted the light curve of the \emph{Chandra} observation, searching for evidence of rotational modulation of the X-ray emission. The 4 light curves for the 4 observations are plotted in Fig.~\ref{fig:lcchandra}. The emission is constant across the 4 segment, showing, similarly to the \xmm\ observation, no evidence of rotationally modulated X-ray emission. No significant flaring is present, and no systematic changes in the X-ray flux are observed between the different observation phases.

\begin{figure*}[htbp]
\begin{center}
\vspace{-2cm}\epsfig{file=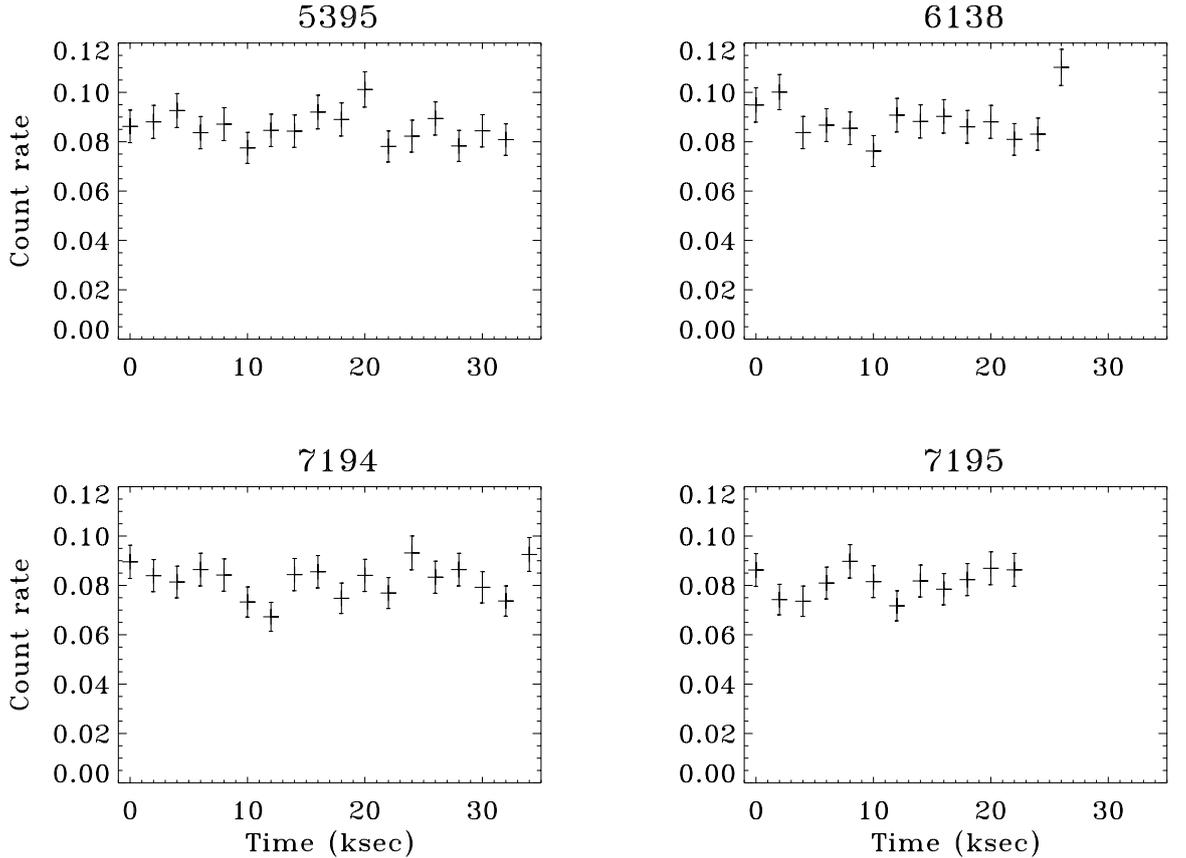, width=20.cm}\vspace{-12.cm}
\caption{The light curve of the 4 \emph{Chandra} observations, integrated at 2 ks binning. No evidence of modulation with the rotational period is present.}
\label{fig:lcchandra}
\end{center}
\end{figure*}

We then searched for evidence of rotational modulation in the radial velocity of the emitting plasma by analyzing both the centroid and the profile of the 19 \AA\ O\,{\sc viii} line. Given the limited phase sampling afforded by the \emph{Chandra} observations we again summed together the two observations close to phase 0.0 and the two observations close to phase 0.25. No evidence of varying Doppler shift with phase is present in our observations: as shown in Table~\ref{tab:o8vels}, the position of the line centroid is constant, with no difference in the nominal position within the error estimate. The same applies to the line's FWHM. 

As mentioned above, the strongest unblended line profile in the dataset is the O\,{\sc viii} resonance doublet near 18.97 \AA.  Gaussian fitting procedures in the data analysis package {\em Sherpa} have been used to measure the centroid positions of the $+1$ and $-1$ order profiles. We find that any velocity shift in the line's centroid is the measurement errors (see Table~\ref{tab:o8vels}). We conclude, therefore, that there is no measurable change in the radial velocity of the X-ray emitting plasma. Given the measurement errors of 0.01 \AA\, we put an upper limit to any relative velocity shifts of $\le 160$ km/s when comparing spectra centered at phases 0.05 and 0.27. Furthermore, the O\,{\sc viii} 18.97 \AA\ line width is consistent with the instrumental broadening, which has a ${\rm FWHM} = 900$ km/s (Table~\ref{tab:o8vels}).

\begin{table} 
\begin{tabular}{lll} 
\hline 
Mean $\phi$ & $\lambda_o$ $(-1)$ [\AA]   &  $\lambda_o$ $(+1)$ [\AA]        \\  \hline 
0.05   & $18.976 \pm 0.010$ &  $18.970 \pm 0.010$   \\ 
0.27    & $18.971 \pm 0.009$ &  $18.977 \pm 0.006$  \\\hline
 &  FWHM $(-1)$ [\AA]        & FWHM $(+1)$ [\AA]       \\  \hline 
0.05    &  $0.07 \pm 0.03$ & $0.06 \pm 0.03$ \\ 
0.27     &  $0.05 \pm 0.01$  & $0.05 \pm 0.02$ \\ 

\end{tabular} 
\label{tab:o8vels} 
\caption{The mean wavelength and FWHM of the O\,{\sc viii} line, grouping observations in two groups of two, maximizing the phase sampling.}
\end{table} 

In order to investigate the level of broadening in the emission line profiles, we summed up the counts from the strongest line profiles in the spectral dataset. The strongest six line profiles all showed peaks with more than 15 counts; these were used to build a composite profile to investigate the origin of the X-ray emission from the system. This profile is built by first measuring the zero-velocity positions of the line profiles using the entire spectrum computed by summing over all four exposures. This is valid as the source position remains within 1 or 2 pixels between all observations; hence the wavelength scale should be consistent between all four pointings. After the zero-velocity positions are established, both the $+1$ and $-1$ order spectra are converted to velocity-space using these zero-velocity offsets; they are then interpolated to the same wavelength resolution (187 km/s or 0.0125 \AA), which corresponds to a line at the mean wavelength (20 \AA) and co-added. The wavelengths of the lines used are centered at: Fe\,{\sc xvii} 15.014 \AA, Fe\,{\sc xvii} 17.0510 \AA, O\,{\sc viii} 18.9725 Fe\,{\sc xvii}, O\,{\sc vii} 21.6015 \AA, O\,{\sc vii} 21.8036 \AA\ and N\,{\sc vi} 24.7792 \AA\ respectively. 

Figure~\ref{fig:chandralineprof} shows the composite profiles for orders $\pm 1$ compared to the combined instrumental profile for the six strongest lines (dashed line). This was produced by scaling the instrumental profile until it fitted the amplitude of the corresponding strong line for each of the six strong line profiles; interpolating the scaled instrumental profile to the same velocity scale used for the composite profile and co-adding all of these scaled, interpolated instrumental profiles to produce a composite 
instrumental profile. A similar analysis performed on phase-grouped spectra also failed to reveal any change in the line profile and positions.

\begin{figure*}[htbp]
\begin{center}
\epsfig{file=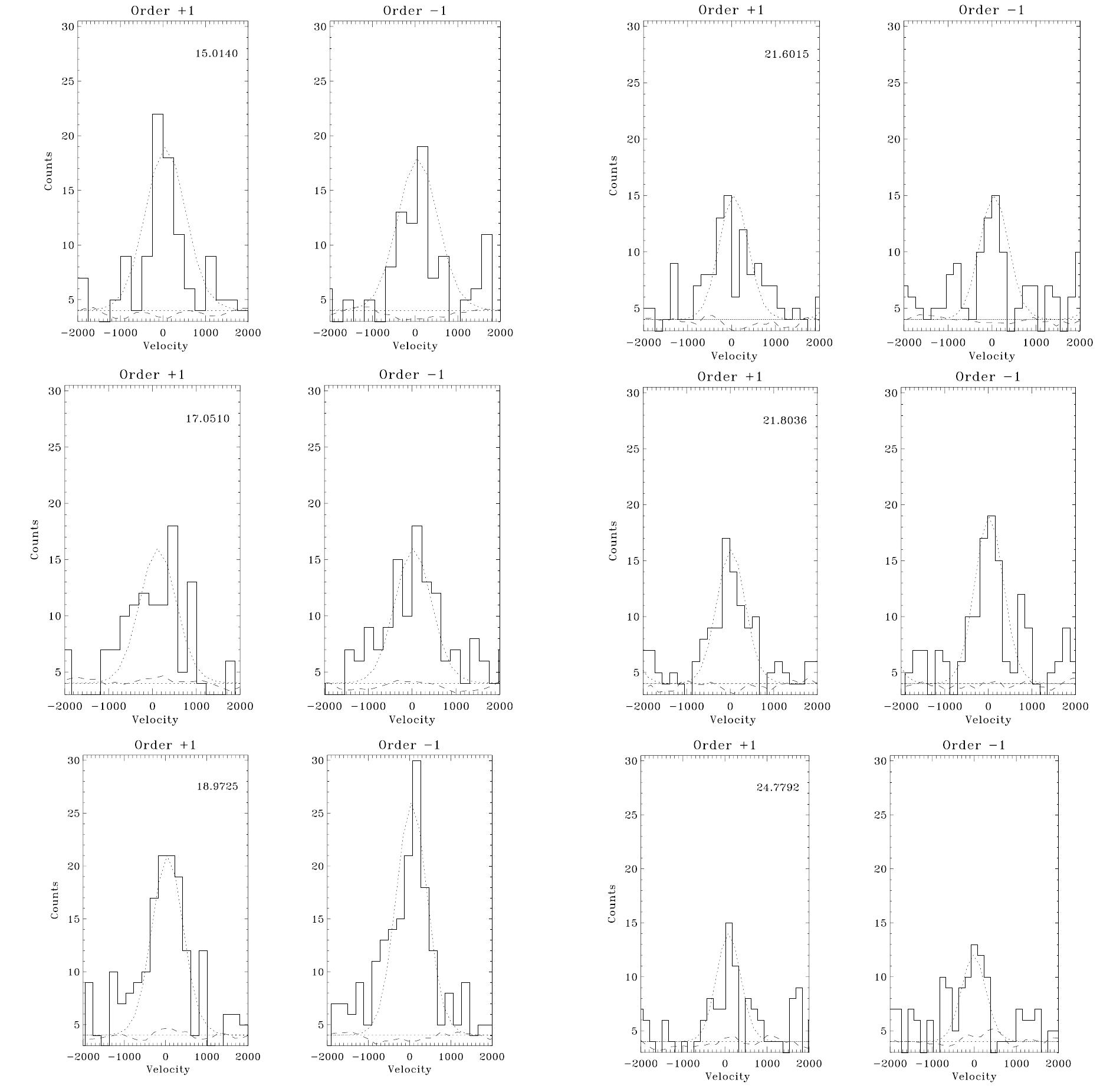,width=16cm}
\caption{The line profiles for the Fe\,{\sc xvii} 15.014 \AA, Fe\,{\sc xvii} 17.0510 \AA, 
O\,{\sc viii} 18.9725 Fe\,{\sc xvii}, O\,{\sc vii} 21.6015 \AA, O\,{\sc vii} 21.8036 \AA\ and N\,{\sc vi} 24.7792\AA\ lines respectively, obtained by adding all 4 \emph{Chandra} observations.}
\label{fig:chandralineprof}
\end{center}
\end{figure*}

\begin{figure}[htbp]
\begin{center}
\vspace{-1.5cm}
\epsfig{file=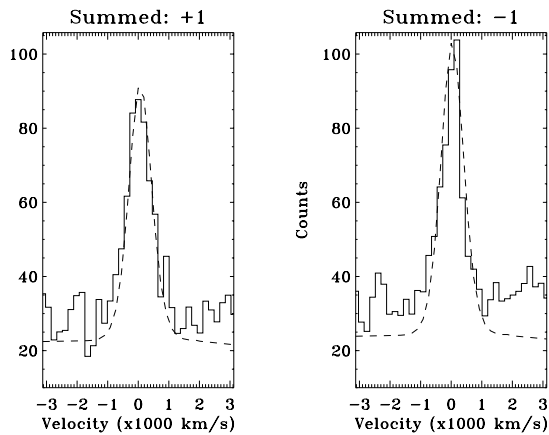,width=10cm,clip=}\vspace{-6.cm}
\caption{The summed line profiles obtained adding up the Fe\,{\sc xvii} 15.014 \AA, Fe\,{\sc xvii} 17.0510 \AA, O\,{\sc viii} 18.9725 \AA, O\,{\sc vii} 21.6015 \AA, O\,{\sc vii} 21.8036 \AA\ and N\,{\sc vi} 24.7792\AA\ lines from the 4 \emph{Chandra} observations. The dotted line shows the composite instrument profile.}
\label{fig:chandrasumlineprof}
\end{center}
\end{figure}

We then analyzed the individual line profiles for evidence of velocity broadening, looking at both the line profiles from the individual observations (Fig.~\ref{fig:chandralineprof}) as well as to the line profile obtained by summing all strong lines (Fe\,{\sc xvii} 15.014 \AA, Fe\,{\sc xvii} 17.0510 \AA, O\,{\sc viii} 18.9725 \AA, O\,{\sc vii} 21.6015 \AA, O\,{\sc vii} 21.8036 \AA\ and N\,{\sc vi} 24.7792\AA) from all 4 observations. This allows (in the hypothesis of no variability from one observation to the other) to obtain a much higher $S/N$ and thus to determine the line profile to a higher accuracy. In fact, all the observed lines are consistent with the instrumental resolution ($\Delta \lambda \simeq 0.05$\,\AA\ for the LETG/HRC instrumental combination), and thus allow us to exclude the presence of plasma moving at velocities $\ga 600$ km s$^{-1}$. In the case of rotational broadening this translates to a maximum distance from the stellar center $d \la 20\,R_{\star}$. While not very constraining, this provides an upper limit to the distance of the X-ray emitting plasma, which, in conjunction with the lower limits provided by the He-like triplet analysis, allows to bound the location of the X-ray emitting region around the star.

The terminal wind velocity of \bcep\ has been estimated between 800 km/s (D01) and 1500 km/s (\citealp{sho+2007}); the lack of observed broadening points toward X-ray emission from a magnetically confined plasma, rather than from a freely flowing wind.

\refcom{We also verified that the $f/i$ ratios in the LETG spectra are consistent with the values derived for the RGS spectra. However, given their lower $S/N$ the LETG $f/i$ ratios have significantly larger associated uncertainties and thus do not help in further constraining e.g.\ the location of the plasma. }

\section{Discussion and conclusions}
\label{sec:concl}

\subsection{X-ray modulation and location of the plasma}

\bcep\ has a close B6-8 companion which could be the component on which the H$\alpha$ emission is located (\citealp{sho+2006}). The companion is very unlikely to be contributing to X-ray emission in the \bcep\ system: in the regime in which X-ray emission is originating in shocks in the wind (i.e. for OB stars) the emission increases with stellar mass and spectral type; therefore the B1 primary is very likely dominating the X-ray emission from the system and thus all our conclusions are unaffected by the presence of the late B-type companion and by the fact that the \bcep\ primary may no longer be considered a Be star.

Our X-ray campaign on \bcep\ shows limited evidence of modulation of its X-ray emission level, both short-term (i.e.\ within a rotational period) and long-term (i.e.\ from one observation to the other), with for example the ROSAT and \xmm\ observations showing very similar flux levels across several years. Evidence of low-level modulation of the X-ray emission with stellar rotation is present in the pn low-resolution X-ray data, while the LETG and RGS high-resolution spectra appear remarkably similar to each other. For instance, as shown in Fig.~\ref{fig:orgs}, the \ovii\ triplet shows no evidence of variation from one rotational phase to the other. For the \emph{Chandra} LETG spectra little if any variability is present in the line fluxes as well as in the centroid of the line positions.

Such high degree of constancy in the X-ray emission is in contradiction with the
high ($\simeq 50\%$) level of rotational modulation expected in the MCWS
framework \refcom{of} D01: with the geometry of \bcep\ the disk is seen alternatively face-one and edge-on. If indeed X-ray emission is due to stationary shocks on either side of the disk (which is expected to have a high optical thickness at soft X-ray wavelengths), during face-on configurations about 50\% of the flux should be absorbed. 

In contrast to what was expected from the MCWS scenario \refcom{by D01}, we find a slightly ($\simeq 10$\%) higher emission level for the face-on configuration with respect to the edge-on configuration. This allows us to exclude the simple scenario proposed by D01 and in particular it definitively rules out the presence of the optically thick layer in the magnetic equator. \refcom{Such smaller modulation, with the observed phase, is along the lines predicted by the dynamical MCWS models of \cite{uo2002} and \cite{goc+2005}, which would indeed predict for the \bcep\ configuration, no thick disk and a modulation of approximately 5\%, fully consistent with the observed values.}

The measurements of the line intensity in the He-like triplets, and in particular the \ovii\ one, allows us to determine the location of the bulk of the emitting plasma. The temperature of formation of the \ovii\ triplet peaks at 0.17 keV, not very different from the characteristic temperature of the bulk of the emitting plasma, as determined from the pn spectra. Therefore, the characteristic distance from the star's photosphere determined from the \ovii\ triplet should be representative of the majority of the plasma at X-ray temperatures. The detection at $\sim$3$\sigma$ of the $f$ line in the \ovii\ triplet allows us to determine a distance range from the photosphere (rather than an upper limit), estimated to be $d \simeq 4\,R_{\star}$ (see sec.~\ref{RGS data}). \refcom{The derived location for the X-ray emitting plasma is also consistent with the location predicted by the \cite{uo2002} and \cite{goc+2005} models, which locate the plasma between the Alfven and Kepler radii, i.e. between $R \simeq 5 \,R_{\star}$ and $R \simeq 7 \,R_{\star}$.}

An additional constraint on the spatial location of the X-ray plasma in \bcep\ comes from the relation between the plasma pressure and the confining magnetic field, under the assumption that the X-ray emitting plasma is indeed magnetically confined, in agreement with the lack of line broadening otherwise expected for a standard wind shock model. The plasma density can be derived from the emission measure, derived either from the global fit to the pn spectrum, or taking the emission in a given line. If we take the flux in the strongest ($r$) \ovii\ line, we derive $E\!M = 1.5 \times 10^{55}$ cm$^{-3}$. Assuming the plasma to be confined in a spherical shell comprised between 4 and $6\,R_{\star}$ (as derived from the triplet ratios), the emitting volume is $V = 7.5\times 10^{37}$ cm$^3$, resulting in an average plasma density $n = 4.5\times 10^8$ cm$^{-3}$. Such low density is consistent with no density effect on the He-like triplets of the elements being considered. The magnetic field needed to confine this plasma (taking into account the average temperature $T \simeq 0.3$ keV) is $B = \sqrt{8\pi 2 n k T} \simeq 3$\,G. Assuming the stellar magnetic field to dipolar, and taking the best estimate for the polar magnetic field of 355 G, and scaling the dipolar field as $d^3$ ($d$ being the distance from the star), the magnetic field at $5\,R_{\star}$ from the photosphere is $355/5^3 \simeq 3$\,G, similar to the pressure required to confine the emitting plasma. This indicates that the emitting plasma is weakly confined by the magnetic field, with $\beta \simeq 1$.

Using the location of the plasma found above ($R_{\rm out} = 6\,R_{\star}$ and $R_{\rm in} = 4\,R_{\star}$) and assuming a geometrically thin disk, we can estimate the variation of the X-ray flux from edge-on to face-on configuration to be of the order of 7\%. This number is compatible within the error bars with the observed variation level. This scenario would rule out the presence of X-ray emitting plasma at high latitudes above the magnetic equator \refcom{and is again compatible with the predictions from the more recent, dynamical MCWS models.}

\subsection{Magnetic confinement}

Standard models of X-ray emission from massive stars, originating in shocks in an unconfined wind, predict both a general blue-shifting of the line together with its broadening. The magnitude of the expected  broadening is comparable with the wind terminal velocity, which in the case of \bcep\ is $\simeq 800$ - $1500$ km/s. The blue-shift depends on the amount of absorption of the line's red wing, and is expected to be less than the broadening. 

The analysis of the LETG line profiles rule out the presence of any broadening above the instrumental line profile, which has a FWHM of approximately 600 km/s, and of blue-shifts greater than 160 km/s. While they are not extremely constraining relative to the wind terminal velocity of \bcep, these elements indicate a lack of any significant bulk motions in the X-ray emitting plasma, and point toward its being magnetically confined, therefore ruling out, for \bcep, X-ray emission from shocks in an unconfined wind.

The relatively low temperature of the X-ray emitting plasma (with the dominant component being at $\approx 3$ MK) together with the lack of significant short-term variability (flares) point however to a lack of magnetic heating (i.e.\ due to magnetic reconnection): in active cool stars, the magnetic reconnection that dominates the heating of the plasma results in  much higher temperatures and stochastic variability. The high temperatures observed for $\theta^1$ Ori C also points to the presence of magnetic heating in some massive stars.

In the case of \bcep\ the magnetic heating only has the apparent function of confining the plasma, which is likely heated to the observed 3 MK by shocks. Also, the bulk of the X-ray plasma appears to be confined by a relatively weak magnetic field, close to the limit where the magnetic pressure becomes too weak to confine the plasma.

\subsection{Comparison with other stars}

The behavior of \bcep\ as observed in our \emph{Chandra} and \xmm\ observations appears to be significantly different from the other well studied example of magnetically confined wind in a massive star, $\theta^1$ Ori C. In that case, the X-ray emission is strongly modulated at the rotational period, by approximately 50\%, as predicted by the MCWS model. At the same time, the observed plasma temperatures are, for $\theta^1$ Ori C, much higher than for \bcep, reaching up to 30 MK. Furthermore, the analysis of the triplets in $\theta^1$ Ori C show that the emitting plasma is located much closer to the star's photosphere than in the case of \bcep. In $\theta^1$ Ori C the bulk of the plasma is located at $\simeq 1.5\,R_{\star}$ while in \bcep\ the cooler plasma traced by the O triplet is located at $\simeq 5\,R_{\star}$, while the hotter plasma traced by the Ne triplet is located at $\la 2\,R_{\star}$ from the photosphere, pointing also to a stratification of the plasma as a function of temperature. Finally, the metal abundances determined for the $\theta^1$ Ori C plasma are much higher than for \bcep. In \bcep\ all elements appear to be depleted with the exception of Si, while many elements are enhanced in the  $\theta^1$ Ori C spectrum, showing a different process in operation.

\cite{goc+2005} have modeled the observed emission from $\theta^1$ Ori C with a dynamic version of the MCWS, showing that both infall to the photosphere and outflow make it unlikely that the thick disk predicted by \cite{bm97a} would actually form. They interpret the observed modulation (which goes in the opposite direction from the one predicted by the D01 model, as also observed by us for \bcep) as occultation of the X-ray emitting plasma (located, in their model, close to the equator) by the stellar photosphere. 

A similar situation, with the dynamics preventing the thick disk from forming, is possibly present in \bcep, so that also in this case the X-ray emission is produced by a magnetically confined wind, with the lower observed temperatures explained by the lower wind velocity.

\subsection{Conclusions}

The \emph{Chandra} and \xmm\ observations discussed here have failed to display the signatures expected by the static MCWS model \refcom{of D01}, in particular the strong rotational modulation of the X-ray emission due to the presence of the cool disk around the star. However the X-ray plasma appears confined, and a small amplitude modulation is visible in the \xmm\ EPIC data. Together with the low temperature and the lack of flaring, this can be interpreted as emission from a magnetically confined wind but without the cool, optically thick disk predicted by the D01 model. \refcom{The observed modulation is fully compatible with the emission scenario predicted by the more recent dynamical MCWS models for a star with the characteristics of \bcep.}

\begin{acknowledgements}

The authors thank the \xmm\ and \emph{Chandra} observatories for their support in the scheduling of a complex phase locked observation. Also the support of L. Scelsi in the analysis of the \xmm\ RGS data is gratefully acknowledged. \refcom{The authors also thank the referee, M. Gagne, for a very detailed and useful report.}

\end{acknowledgements}

%\bibliographystyle{aa}
%\bibliography{/Users/Fabio/Documents/Bibliografia/references}

\end{document}